\documentclass[twocolumn]{aastex631}

\usepackage{graphicx}
\usepackage{amsmath}
\usepackage{amssymb}
\usepackage{makecell}
\usepackage[table]{xcolor}

\def\H{\hbox{{\rm H~}}}
\def\HI{\hbox{{\rm H}{\sc \,i}}}
\def\HeI{\hbox{{\rm He~}\kern 0.1em{\sc i}}}
\def\HII{\hbox{{\rm H~}\kern 0.1em{\sc ii}}}
\def\Lya{\hbox{{\rm Ly}\kern 0.1em$\alpha$}}
\def\Lyb{\hbox{{\rm Ly}\kern 0.1em$\beta$}}
\def\Lyg{\hbox{{\rm Ly}\kern 0.1em$\gamma$}}
\def\Lyth{\hbox{{\rm Ly}\kern 0.1em$\theta$}}
\def\Lyfive{\hbox{{\rm Ly}\kern 0.1em$5$}}
\def\Lysix{\hbox{{\rm Ly}\kern 0.1em$6$}}
\def\Lyseven{\hbox{{\rm Ly}\kern 0.1em$7$}}
\def\Lyeight{\hbox{{\rm Ly}\kern 0.1em$8$}}
\def\Lynine{\hbox{{\rm Ly}\kern 0.1em$9$}}
\def\Lyten{\hbox{{\rm Ly}\kern 0.1em$10$}}
\def\kms{\hbox{km~s$^{-1}$}}
\def\cmsq{\hbox{cm$^{-2}$}}
\def\cc{\hbox{cm$^{-3}$}}

\definecolor{myblue}{HTML}{1f77b4}
\definecolor{myred}{HTML}{d62728}
\definecolor{mygray}{HTML}{808080}
\definecolor{lightpink}{RGB}{255, 228, 225}

\newcommand{\galpak}{GaLPaK$^\text{3D}$}

\shortauthors{S. R. Bharadwaj et al.}
\shorttitle{pLLS in a Galaxy-Pair Halo: A Possible Signature of Gas Inflow}
\begin{document}

\title{A Partial Lyman Limit Absorber in the Halo of a Galaxy Pair: A Possible Signature of Gas Inflow}

\author[0009-0003-1664-3369]{Sumukha R. Bharadwaj}
\affiliation{Department of Earth and Space Sciences, Indian Institute of Space Science and Technology, Thiruvananthapuram 695547, India}
\email{sumukharb.24@res.iist.ac.in}

\author[0000-0002-1490-5367]{Anand Narayanan}
\affiliation{Department of Earth and Space Sciences, Indian Institute of Space Science and Technology, Thiruvananthapuram 695547, India}

\author[0000-0003-3938-8762]{Sowgat Muzahid}
\affiliation{Inter-University Centre for Astronomy and Astrophysics (IUCAA), Post Bag 4, Ganeshkhind, Pune 411007, India}

\author[0000-0003-4877-9116]{Jane C. Charlton}
\affiliation{Department of Astronomy and Astrophysics, The Pennsylvania State University, University Park, PA 16802, US}

\author[0000-0001-5804-1428]{Sebastiano Cantalupo}
\affiliation{Astrophysics Unit, Department of Physics, University of Milano-Bicocca, Piazza della Scienza 3, 20126 Milan, Italy.}


\begin{abstract}
We present an analysis of a partial Lyman limit system at $z = 0.87641$ detected in the $HST$/COS spectrum of the background quasar LBQS~0107$-$0235. The absorber exhibits a simple kinematic structure, with the metal-lines and the {\ion{H}{1}} Lyman-series absorption well described by a single component. Photoionization modeling yields a gas metallicity of one-tenth solar and a hydrogen number density of $n_\text{H} \approx 8.5 \times 10^{-4}$~{\cc} ($\log_{10}(n_\text{H}/\mathrm{cm}^{-3}) \approx -3.1$). At the absorber redshift, the $VLT$/MUSE data show two galaxies (G1 and G2) at normalized impact parameters of $\rho / R_\mathrm{vir} \approx 0.9$ and velocity separations of $|\Delta v| = 18$ and $99$~{\kms}, respectively, from the absorber. Both galaxies have rotating disks with stellar masses of $M_\ast \approx 6\times10^{9}$ and $\approx 2.2 \times10^{10}$~M${_\odot}$. Their 100-Myr-averaged star formation rates are $\approx 2.5$ and $\approx2.2$~M$_\odot~\mathrm{yr}^{-1}$, though their instantaneous rates place G2 on the star-forming main sequence and G1 above it, which is actively star-forming at this redshift. The absorber is positioned very close to the projected major axis of both galaxies. The absorber's orientation, kinematics, and sub-solar metallicity ($\log_{10}(Z/Z_\odot) = -1.05$) are consistent with the absorption tracing a sub-solar metallicity inflowing stream, though a galaxy--galaxy interaction origin cannot be excluded. We discuss these scenarios in the context of cosmological simulations of cold-mode accretion and CGM gas flows around galaxies with halos of mass $M_\text{h} \lesssim 10^{12}$~M$_\odot$.
\end{abstract}


\section{INTRODUCTION}
\label{sec:introduction}

Our understanding of galaxy evolution is predicated on the cosmic baryon cycle, a continuous and complex interplay of gas flows in and out of galaxies. Galaxies are not isolated islands but are deeply embedded within vast, diffuse gaseous envelopes known as the circumgalactic medium (CGM), which extends from the galactic disk to the virial radius of the host dark matter halo. This medium serves as the critical interface between a galaxy and the wider intergalactic medium (IGM). It acts both as the fuel reservoir for star formation and as the repository for metal-enriched material ejected by galactic winds, which can later be re-accreted by the galaxy. 
Due to its extremely low density, detecting the CGM in emission is challenging. The most effective technique for studying this diffuse gas is through absorption-line spectroscopy, using the bright, continuous spectra of background quasars as probes. This method has revealed that the CGM is a dynamic multiphase medium, characterized by a wide range of physical conditions, chemical compositions, and gas kinematics.

Quasar absorption line studies have established a fundamental framework for characterizing the CGM, in which the neutral hydrogen column density, $N(\mathrm{H\,\textsc{1}})$, provides a first-order classification of absorbing gas. While the low-density Ly$\alpha$ forest traces the IGM, the transition toward the gas-rich environments of galaxies is marked by a continuous distribution of absorbers with progressively higher column densities. This hierarchy spans from strong Ly$\alpha$ forest systems (SLFSs) and partial/Lyman limit systems (pLLSs/LLSs) to the high-column density damped Ly$\alpha$ absorbers (DLAs) \citep{Steidel1992, Rao2000, Lehner2022, Dutta2025}. In contrast to DLAs, which predominantly trace the inner ISM and galactic disks, LLSs and pLLSs serve as unique probes of the CGM and the transitional interfaces between the halo and the IGM \citep{Cooper2015, Fumagalli2016, Lehner2022}. The physical connection between these absorbers and their host halos is substantiated by a strong inverse correlation between absorption strength and projected impact parameter. This trend--robustly established by large-scale surveys such as \texttt{MUSEQuBES}, \texttt{MAGG}, \texttt{MEGAFLOW}, and \texttt{CUBS}--holds for both \ion{H}{1} and common metal ions, confirming that these systems trace the multiphase circumgalactic environment \citep{Chen2000, Chen2010, Werk2014, Prochaska2017, Tchernyshyov2022, Dutta2024, Weng23}. The baryonic overdensities inferred for these systems ($\delta \sim 10^2$--$10^3$) are consistent with gas that is gravitationally bound to galactic halos rather than the diffuse IGM \citep{Wotta_2016, Tumlinson2017}.

High-resolution absorption-line studies show the CGM to be multiphase, each traced by distinct ionic species. The cool, photoionized phase ($T~\approx~10^4~$K) is traced by low-ionization species such as \ion{H}{1}, \ion{C}{2}, \ion{Si}{2}, and \ion{Mg}{2}. This phase constitutes the bulk of the neutral and low-ionization gas, tracing diverse structures including near-pristine inflowing streams \citep{Ribaudo2011, Fumagalli2011}, metal-enriched outflows \citep{Bouche12, Schroetter2019}, and extended, co-rotating gaseous disks \citep{Ho17, Zabl2019, Weng23, Udhwani_2025}. In contrast, a warm-hot, collisionally ionized phase ($T~\approx~10^5~-~10^6~$K) is traced by high-ionization species such as \ion{O}{6} and \ion{Ne}{8} \citep{Narayanan10, Narayanan2018, Meiring2013}. In this context, observations show that \ion{O}{6} absorption is commonly detected around star-forming galaxies, with high covering fractions extending to large projected distances from galaxies \citep{Tchernyshyov2022, Mishra2024, Dutta2025, Qu2024}. The frequent co-existence of low- and high-ionization species within the same absorption systems, often with related but non-identical kinematics, points to complex physical processes operating at the interfaces between gas phases \citep{Fox2006, Savage2010, Savage2012, Narayanan_2018, Khonde2024}.

Observational surveys also find a positive correlation between metallicity and $N(\text{\ion{H}{1}})$, suggesting that the higher-density gas is systematically more chemically enriched \citep{Wotta_2016, Wotta2019, Lehner2022}. However, the absence of a monotonic radial metallicity gradient also implies that the CGM is poorly mixed, containing gas with diverse enrichment histories \citep{Ribaudo2011, Lehner2013, Quiret2016, Prochaska2017, Lehner2019, Sankar20}. This chemical complexity is mirrored by the gas kinematics, typically characterized through integral field unit (IFU) spectroscopy of the absorber host galaxies. Comparisons of the galaxy kinematic maps derived from the IFU data with the specific kinematics of the absorbing gas reveal an azimuthal anisotropy in the CGM, where metal-enriched outflows preferentially align with the galaxy's minor axis, whereas the near-pristine accretion flows are found co-rotating along the major axis \citep{Bordoloi11, Bouche2011, Kacprzak_2012, Schroetter2019, Zabl2019}. 

The IFU observations also help in connecting the physical and dynamical properties of the absorbing gas to the properties and environments of their host galaxies \citep{Bouche2013, Peroux14, Weng23}. Observational campaigns such as the \texttt{MEGAFLOW} \citep{Schroetter2016, Zabl2019, Bouche2025}, \texttt{MAGG} \citep{Lofthouse2020, RDutta2020, Galbiati2023}, \texttt{MUSEQuBES} \citep{Muzahid2020, Dutta2024, Banerjee2025}, $\text{CGM}^2$ \citep{Wilde2021, Tchernyshyov2022}, and \texttt{CUBS} surveys \citep{Chen2020, Cooper2021, Zahedy21, Qu2024, Mishra2024} have established that the CGM is a complex ecosystem sensitive to galaxy stellar mass \citep{Tchernyshyov2022, Dutta2025}, star-formation history \citep{Tchernyshyov2023, Langan2023}, and local environment \citep{Hamanowicz20, RDutta2020, Cherrey2024, Dutta2025}. Many absorption systems arise in environments containing multiple galaxies rather than a single isolated host. In such cases, identification of the gas origin necessitates a joint consideration of the morpho-kinematics of all galaxies in the vicinity rather than solely the nearest neighbor \citep{Peroux19, Hamanowicz20, Kulkarni22, Khonde2024}. Recent analyses from the MUSE-ALMA halos survey demonstrate that strong absorbers at $z \lesssim 1.0$ frequently trace complex environments containing multiple galaxies within impact parameters of $\approx 50$ kpc. In such systems, distinguishing gas accretion from tidal streams or the intragroup medium requires precise constraints on the kinematic coupling between the absorbing gas and the morpho-kinematics of the potential host galaxies \citep{Hamanowicz20, Peroux2022, Weng23}. Such scenarios highlight the importance of considering the "group-centric" or shared halo environment in interpreting absorption systems. 
This work aims to contribute to this effort by presenting a detailed analysis of a multiphase absorption system and its connection to a proximate pair of star-forming galaxies, and their overlapping CGM. 

In this work, we investigate a partial Lyman limit system (pLLS) at $z = 0.87641$ detected in the spectrum of the background quasar LBQS~0107$-$0235. Using far-ultraviolet absorption-line data from $HST$/COS and integral-field spectroscopic observations from $VLT$/MUSE, we characterize the properties of the absorbing gas and examine its connection to nearby galaxies. The structure of this paper is as follows. Section~\ref{sec:observations} contains details on the \textit{HST}/COS and $VLT$/MUSE data. Section~\ref{sec:analysis_of_absorption} presents the analysis of the absorption system, including column density measurements and kinematic structure. In Section~\ref{sec:ionization_modeling}, we use photoionization modeling to infer the physical conditions and chemical composition of the absorbing gas. Section~\ref{sec:associated_galaxies} investigates the galaxies associated with the absorber, including their morpho-kinematic properties, star formation activity, and stellar populations, and explores whether they are capable of driving galactic-scale winds. In Section~\ref{sec:origin_of_absorption}, we assess the origin of the absorbing gas in the context of galaxy-halo interactions and gas accretion scenarios. Finally, Section~\ref{sec:summary} summarizes our main findings. Throughout this work, we adopt a flat $\Lambda$CDM cosmology with $H_0 = 67.66\,\mathrm{km\,s^{-1}\,Mpc^{-1}}$, $\Omega_{\rm m} = 0.3111$, and $\Omega_\Lambda = 0.6889$ \citep{Planck_Collab_2020}.

\section{OBSERVATIONS}
\label{sec:observations}

\subsection{$HST$/COS Spectroscopic Data}
\label{sec:cos_data}
The far-ultraviolet (FUV) spectrum of the background quasar {\mbox{LBQS~0107$-$0235}} (RA = 01$^{\rm h}$10$^{\rm m}$13.160$^{\rm s}$, Dec = $-02^{\circ}19^{\prime}52.84^{\prime\prime}$; $z_{\rm em} = 0.960$) was acquired with the Cosmic Origins Spectrograph \citep[COS;][]{Green2012} on board the \textit{Hubble Space Telescope} as part of Program ID 11585 (PI: N. Crighton). A total of 17.05 ks was spent with the G130M grating and 19.95 ks with the G160M grating, providing continuous spectral coverage over 1130--1800 \AA\ at a resolving power of $R \simeq 20{,}000$ ($\approx 15-18$ km s$^{-1}$ FWHM). The coadded spectrum was retrieved from the $HST$ Spectroscopic Legacy Archive \citep[HSLA;][]{peeples}. The $HST$/COS data presented in this article were obtained from the Mikulski Archive for Space Telescopes (MAST) at the Space Telescope Science Institute. The COS spectra, which are oversampled, were rebinned to two pixels per resolution element (channel width $\Delta\lambda \approx 0.06$ \AA) to improve the per-pixel signal-to-noise ratio. The resulting spectrum reaches a median S/N $\approx 12$ per resolution element in continuum regions. Low-order polynomials were fitted to line-free windows of $\approx 20$~{\AA} to normalize the continuum.

\subsection{$VLT$/MUSE IFU Data}
\label{sec:ifu_data}

The quasar field was observed using the $VLT$/Multi-Unit Spectroscopic Explorer \citep[MUSE;][]{Bacon2010} integral field spectrograph in its wide-field mode, covering a $1^{\prime} \times 1^{\prime}$ field of view. The data cube spans a spectral range of 4750--9350 \AA\ with a spectral resolution varying from $R \approx 1800$ to $R \approx 3500$. The spatial sampling is $0.2^{\prime\prime}$ per pixel. The seeing during the observations is estimated to be $0.60^{\prime\prime}$ using PampelMuse \citep{Kamann2013}. The data were reduced using a combination of the standard ESO MUSE pipeline v1.6 and the \texttt{CubExtractor} package \citep[\texttt{CubEx};][]{cubex}. The \texttt{CubEx} routine \texttt{CUBEFIX} was used to perform a self-calibration on the sky background to correct for flat-fielding residuals. Following that, the \texttt{CUBESHARP} routine was utilized to further suppress sky residuals through a local, flux-conserving sky subtraction that accounts for spatial variations in the instrument's line spread function \citep{cubex}. To isolate faint galaxies, a subtraction of the sky continuum and the bright quasar PSF was necessary. This was accomplished using the \texttt{CUBEBKGSUB} and \texttt{CUBEPSFSUB} routines within \texttt{CubEx}, which employ median filtering and an empirical PSF reconstruction method, respectively. Following the PSF subtraction, to identify galaxy counterparts, we performed source detection to identify galaxy counterparts. We used \texttt{SExtractor} \citep{Bertin1996} on a quasar PSF subtracted white-light image generated from the MUSE data cube, adopting a detection threshold of $1.5\sigma$ above the background and a minimum area of 5 pixels. This process yielded an initial catalog of 121 sources. We filtered this catalog to isolate extended objects using the \texttt{CLASS\_STAR} parameter, resulting in a final sample of 80 galaxy candidates. From the resulting catalog, G1 and G2 were identified as the primary counterparts based on their spatial and kinematic proximity to the $z=0.87641$ absorber. One-dimensional spectra for these candidates were then extracted using the MUSE Python Data Analysis Framework \citep[\texttt{MPDAF};][]{mpdaf}, with apertures defined by the \texttt{SExtractor}-generated segmentation map. Initial redshifts were estimated using the \texttt{MARZ} package \citep{marz}. To achieve the precision required for a detailed kinematic analysis, we performed Gaussian fitting of the H$\beta$ and [\ion{O}{2}] $\lambda\lambda 3726, 3729$ emission lines using \texttt{MPDAF} \citep{mpdaf}. The final redshifts were determined from a weighted average of these line centroids, yielding $z = 0.87630 \pm 0.00003$ for G1 and $z = 0.87579 \pm 0.00005$ for G2. The specific search radius and velocity criteria used to confirm these associations are detailed in Section~\ref{sec:associated_galaxies}. Finally, to accurately model the kinematics of these closely separated galaxies, a mutual PSF subtraction was performed, as detailed in Section~\ref{sec:galpak_analysis}.

\begin{figure*}
    \centering
    \includegraphics[width=\textwidth, height=0.92\textheight, keepaspectratio]{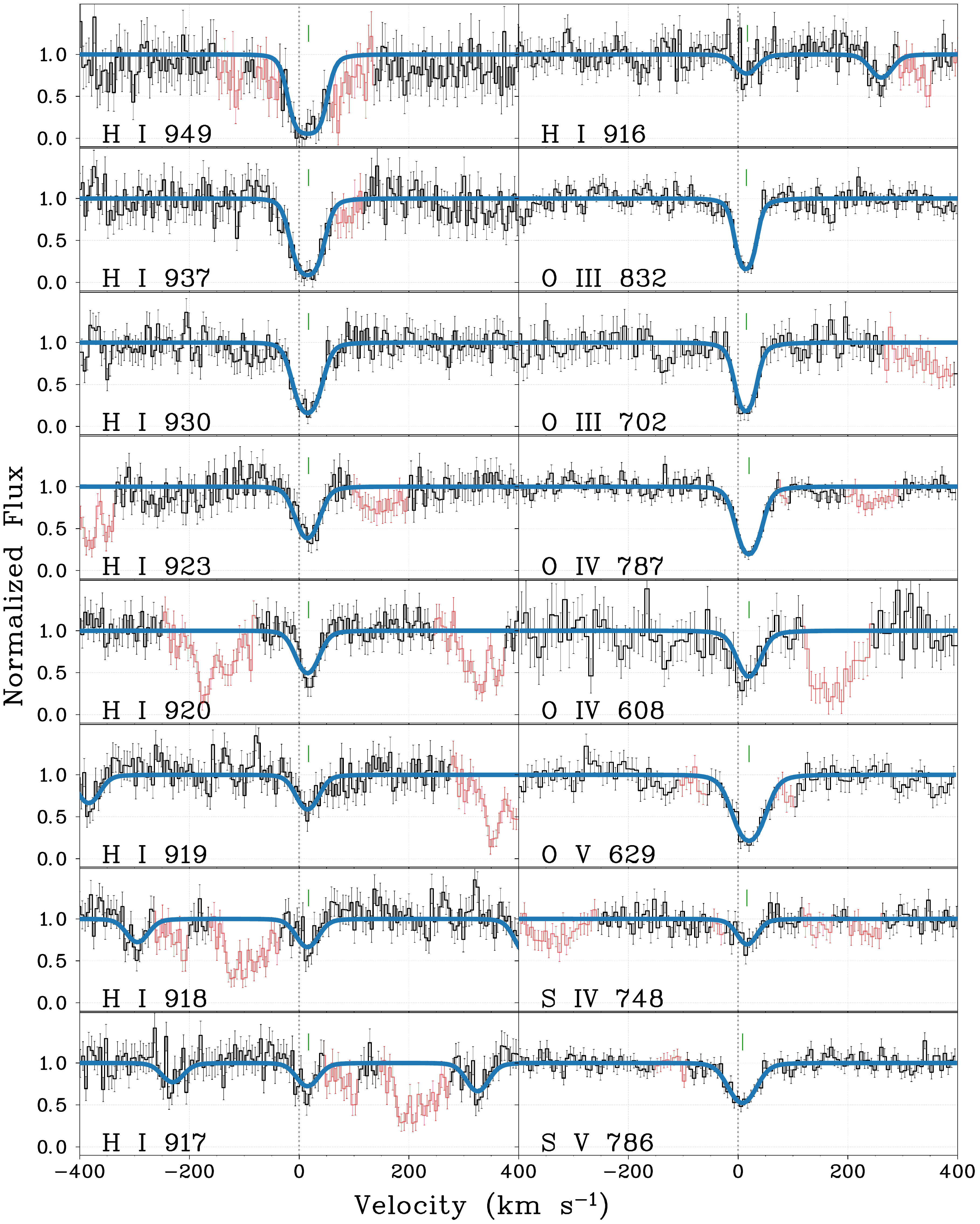}
    \caption{Velocity plot of the $z=0.87641$ absorption system. The normalized $HST$/COS flux is shown in black. The velocity x-axis is centered at $v=0$ \kms, defined by the systemic redshift of galaxy G1 ($z = 0.87630$, \textit{vertical dotted line}). The systemic redshift of galaxy G2 ($z=0.87579$) is at $\approx-81$ \kms. The best-fit single-component Voigt profile (convolved with the COS LSF) for the main absorber at $v \approx 18$ \kms\ (\textit{green vertical markers}) with respect to the redshift of G1 is overlaid in \textcolor{myblue}{blue}. Panels show \ion{H}{1} Lyman series (949 \AA\ to 917 \AA), \ion{O}{3}, \ion{O}{4}, \ion{O}{5}, \ion{S}{4}, and \ion{S}{5} lines. Absorption lines from unrelated systems, treated as contamination, are marked in \textcolor{myred}{red}. The \ion{S}{5} centroid is offset by $\approx - 10$~{\kms} from the \ion{H}{1} and metal lines. There is a possibility of this feature being (or contaminated by) {\Lya} from $z = 0.21387$ as reported by \citet{Tejos}.}
    \label{fig:system_plot}
\end{figure*}

\section{ANALYSIS OF THE ABSORPTION SYSTEM}
\label{sec:analysis_of_absorption}

Figure \ref{fig:system_plot} presents the absorption profiles of the ions detected at $z \approx 0.8764$ in the COS spectrum of the quasar LBQS 0107-0235. The full system plot, including prominent non-detections, is shown in Appendix Figure \ref{fig:sys_plot_page1}. A visual inspection of the spectral features shows that the absorption is kinematically simple. The higher-order \ion{H}{1} Lyman series lines, along with the transitions from the three oxygen ionization stages (\ion{O}{3}, \ion{O}{4}, and \ion{O}{5}), \ion{S}{4}, and \ion{S}{5} are dominated by a single relatively narrow component. This simple kinematic structure motivates modeling the system with a single-component Voigt profile.

The column density ($N$), Doppler $b-$parameter, and velocity centroids of the absorption lines ($v$) were determined by fitting Voigt profiles using the Python package \texttt{VoigtFit} v0.21.7 \citep{voigtfit}. To accurately model the instrumental broadening, the theoretical profiles were convolved with the appropriate $HST$/COS Line Spread Function (LSF)\footnote{\url{https://www.stsci.edu/hst/instrumentation/cos/performance/spectral-resolution}} for the corresponding grating and lifetime position \citep{Ghavamian2009}. To break potential degeneracies between combinations of $N$ and $b$ resulting in similar fits, absorption profiles of multiple transitions of a given species were fitted simultaneously.

For the \ion{H}{1}, the absorber's redshift of $z \approx 0.8764$ places the Ly$\alpha$, Ly$\beta$, and Ly$\gamma$ at wavelengths beyond the coverage of the COS FUV gratings. Our fitting procedure therefore, relies exclusively on the higher-order Lyman series transitions (from \ion{H}{1}~ 949 \,\AA\ and above) to constrain \ion{H}{1}. The simultaneous fitting of the multiple \ion{H}{1} lines resulted in a single component centered at $v = 17~ \pm~ 1$ km s$^{-1}$ relative to the redshift of the galaxy G1, with a column density of $\log_{10}[N(\text{\ion{H}{1}})/\text{cm}^{-2}] = 16.03 \pm 0.04$ and a Doppler parameter of $b = 29 \pm 2$ km s$^{-1}$. The estimate is consistent with the absence of a partial or full Lyman break in the COS spectrum at the redshifted location of the Lyman limit. The measured \HI\ Doppler parameter sets a purely thermal upper limit on the gas temperature of $T_\mathrm{max} = m_p b^2 / 2k_\mathrm{B} \approx 5 \times 10^4$~K, consistent with the assumption of photoionization equilibrium for the low-ionization absorbing gas. While the stronger \HI\ transitions are well reproduced by the profile fits, the higher-order Lyman lines appear somewhat narrower than the model, suggesting the possibility of unresolved substructure. We explored two-component fits for \HI. However, at the spectral resolution and signal-to-noise of the COS data, such models do not yield well-constrained solutions. In the absence of clear kinematic substructure, we adopt a single-component fit as a conservative description of the absorption. The metal lines also show no evidence for multiple distinct components. In any case, the different fitting approaches do not significantly affect the total \HI\ column density used in the modelling.

At the COS resolution, a single component was sufficient to explain the lines of \ion{O}{3}, \ion{O}{4}, \ion{O}{5}, \ion{S}{4}, and \ion{S}{5} as well. The final parameters for all fitted species are presented in Table~\ref{tab:voigt_fit_results}. The velocity centroids of the metal lines and {\ion{H}{1}} agree within $\approx 5$~{\kms}, except {\ion{S}{5}~786}, whose centroid is offset from the {\ion{H}{1}} by $\approx 10$~{\kms}, which is within one resolution element of COS. \citet{Tejos} identify the absorption feature we attribute to {\ion{S}{5}} as {\Lya} at $z = 0.21387$ with $\log_{10} [N(\text{\ion{H}{1}})/\cmsq] = 13.46$ and $b = 25$~{\kms}. However, no higher-order Lyman series line or metal lines are detected at that redshift to support this interpretation, which is expected given the weakness of the feature if it is {\Lya}.

Despite the remarkable alignment of the oxygen ion velocities, the \ion{O}{3} absorption line is narrower than the \ion{O}{4} and \ion{O}{5} lines, as shown by their respective $b$ parameters. This could result from multiple density-temperature phases along the line of sight through the absorbing medium, each contributing to the observed absorption. However, the absorption profiles themselves show no evidence for multiple, well-resolved components, indicating that any such structure must be kinematically blended at the COS resolution. This is further discussed in Section \ref{sec:ionization_modeling}.

We also measured column densities using the apparent optical depth (AOD) method \citep{Savage_1991}. This technique provides a model-independent measurement of column densities for unsaturated lines and also helps to identify unresolved saturation when multiple lines from the same ionic species are involved. The equivalent widths and the column densities derived from the AOD method are presented in Table~\ref{tab:absorption_lines_updated}. The {\ion{O}{4}}~608, 787 lines show signs of unresolved saturation, as indicated by the higher AOD column density measured from the 608~\AA\ line relative to the 787~\AA\ line by $0.16$~dex. We adopt the Voigt-profile fitting results for all ions uniformly, including {\ion{O}{4}}, as the simultaneous fit to both transitions convolved with the COS LSF is more robust to mild saturation than the per-line AOD integration. The integrated AOD column density of {\ion{O}{5}}~629 is $\sim 0.1$~dex lower than the value from profile fitting. While this difference is suggestive of mild unresolved saturation, it is also comparable to the measurement and systematic uncertainties. We adopt the profile-fitting result for the ionization modeling, as it more robustly accounts for the instrumental line spread function in the presence of potential broadening or mild saturation. The agreement between the AOD and Voigt-fit measurements for \ion{O}{3}, and \ion{S}{4} suggests that these lines are free from unresolved saturation effects.

\begin{deluxetable}{lccc}
\tablecaption{The velocity centroids, Doppler $b-$parameters, and column densities for ions modeled with Voigt profiles. Velocities are relative to $z_{G1} = 0.87630$. The listed parameters represent the best-fit single-component model for each species.}
\label{tab:voigt_fit_results}
\tablehead{
\colhead{Line} & 
\colhead{$v$ (km s$^{-1}$)} & 
\colhead{$b$ (km s$^{-1}$)} & 
\colhead{$\log_{10}(N/\mathrm{cm}^{-2})$}
}
\startdata
\makecell[l]{\ion{H}{1} 949/937/930/923/\\920/919/918/917/916} & $17\pm1$ & $29\pm2$ & $16.03\pm0.04$ \\
\ion{O}{3} 702/832 & $15\pm1$ & $13\pm3$ & $14.86\pm0.32$ \\
\ion{O}{4} 608/787 & $20\pm1$ & $21\pm2$ & $14.69\pm0.06$ \\
\ion{O}{5} 629 & $20\pm2$ & $30\pm3$ & $14.14\pm0.05$ \\
\ion{S}{4} 748 & $16\pm4$ & $21\pm7$ & $13.26\pm0.10$ \\
\ion{S}{5} 786$^{a}$ & $8\pm2$ & $28\pm3$ & $13.14\pm0.03$ \\
\enddata
\tablenotetext{a}{The \ion{S}{5} absorption feature is likely contaminated by Ly$\alpha$ at $z = 0.21387$ \citep{Tejos}, which explains the observed velocity offset. Consequently, this measurement is treated as an upper limit in the photoionization modeling (see Section \ref{sec:ionization_modeling}).}
\end{deluxetable}

\section{IONIZATION MODELING OF THE ABSORBING GAS}
\label{sec:ionization_modeling}
To infer the properties of the absorbing gas, we performed photoionization modeling using the code \textsc{cloudy} v23.00 \citep{cloudy}. We employed a Bayesian Markov Chain Monte Carlo (MCMC) framework to constrain the model parameters based on the observed ionic column densities. We utilized \textsc{cloudy} to generate a grid of models assuming photoionization equilibrium (PIE) for a plane-parallel gas slab of constant hydrogen density ($n_\text{H}$) and metallicity, maintained in a steady state of thermal and ionization balance with the extragalactic UV background (UVB) \citep{Fumagalli2016, Zahedy21, Banerjee2025}. The models were computed at the absorber redshift, $z = 0.87641$, using a standard plane-parallel geometry \citep{Lehner2019, Cooper2021}. For UVB, we adopt the model of \citet{Khaire19}, which incorporates updated emissivities from both quasars and star-forming galaxies.  Given the single-component nature of the absorption profiles and their close alignment in velocity space, the ionization models are based on the assumption that the ions arise from a single, co-spatial medium in photoionization equilibrium (PIE). This assumption is further supported by the non-detection of \ion{S}{6} and \ion{Ne}{8} (see Fig \ref{fig:sys_plot_page4}), both of which fall within the COS wavelength coverage at this redshift; their absence is consistent with the gas being photoionized rather than collisionally ionized. The derived properties should therefore be interpreted as representative of the medium as a whole, without accounting for potential small-scale variations.

To explore the parameter space, we generated a grid of \textsc{cloudy} models. The primary input parameters for the grid were the hydrogen number density ($\log_{10}(n_\text{H}/\mathrm{cm}^{-3})$) varied from $-5$ to $-1$ in increments of $0.02$~dex and the gas-phase metallicity ($\log_{10}(Z/Z_{\odot})$) sampled across a range of $-3$ to $+1$. The models assume the solar abundance pattern from \citet{Grevesse}. The models were computed for a reference \ion{H}{1} column density of $\log_{10}[N(\text{\ion{H}{1}})/\text{cm}^{-2}] = 16.03$. The \texttt{emcee} Python package \citep{emcee_2013} was used to perform an MCMC analysis, as described in \citet{Acharya22}. The MCMC samples three primary parameters: the hydrogen number density ($\log_{10} n_\text{H}$), the gas-phase metallicity ($\log_{10} Z/Z_{\odot}$), and the \ion{H}{1} column density ($\log_{10} N(\text{\ion{H}{1}})$). The log-likelihood function for ionic detections is defined as: 
\begin{equation*} 
\ln \mathcal{L} = -\frac{1}{2} \sum_{1} \left[ \frac{(N_{\mathrm{obs},i} - N_{\mathrm{model},i})^2}{\sigma_{i}^2} + \ln(2\pi\sigma_{i}^2) \right] \end{equation*} 
where $N_{\mathrm{obs},i}$ and $\sigma_{i}$ are the observed column densities and associated $1\sigma$ uncertainties listed in Table~\ref{tab:voigt_fit_results}, and $N_{\mathrm{model},i}$ are the values predicted by \textsc{cloudy} \citep{Fumagalli2016, Cooper2015}. Given the $\Delta v \approx 10$~{\kms} velocity offset of \ion{S}{5} relative to the other ions, and the possibility of the line being contaminated by {\Lya} at $z \approx 0.213$ \citep{Tejos}, we consider the observed \ion{S}{5} column density as an upper limit rather than a detection. In our Bayesian framework, this is implemented as a one-sided penalty: the log-likelihood contribution from \ion{S}{5} is zero when the model-predicted column density does not exceed the observed upper limit ($N_{\rm model} \leq N_{\rm obs}$), and incurs a Gaussian penalty of $-\frac{1}{2}\left[(N_{\rm model} - N_{\rm obs})/\sigma\right]^2$ when it does ($N_{\rm model} > N_{\rm obs}$), thereby disfavouring solutions that overproduce \ion{S}{5} while placing no constraint on solutions below the limit. We adopted uniform priors for the density and metallicity, while a Gaussian prior was applied to the \ion{H}{1} column density, centered at the observed value of $16.03$ with a width of $\sigma = 0.04$~dex, corresponding to the measurement uncertainty. The MCMC analysis was performed using 512 walkers for 10,000 steps, following a 5,000-step burn-in phase to ensure convergence.

\begin{figure*}
\centering

\begin{minipage}[t]{0.42\textwidth}
    \centering
    \includegraphics[width=\linewidth]{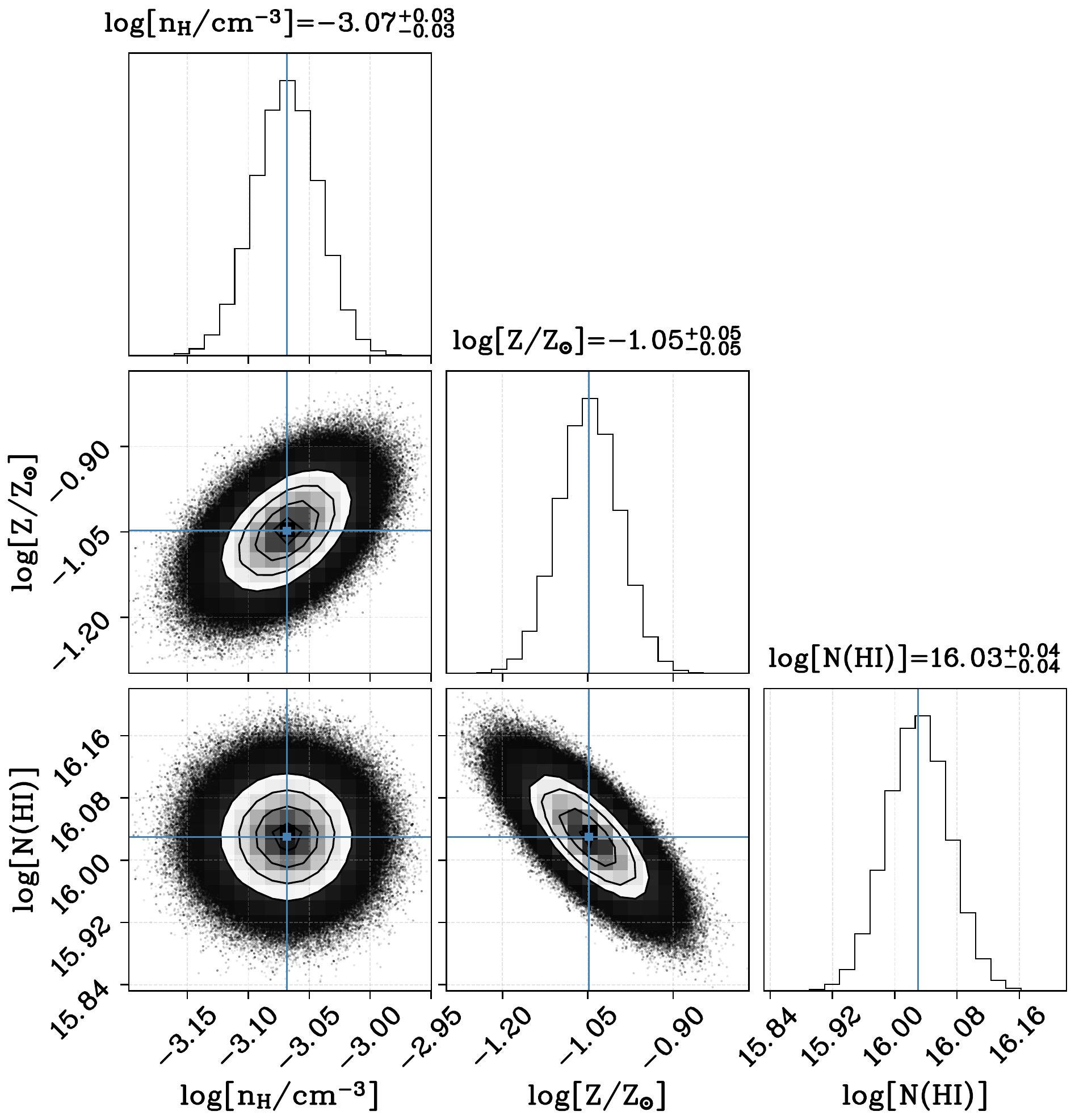}
\end{minipage}
\hfill
\begin{minipage}[t]{0.56\textwidth}
    \centering
    \includegraphics[width=\linewidth]{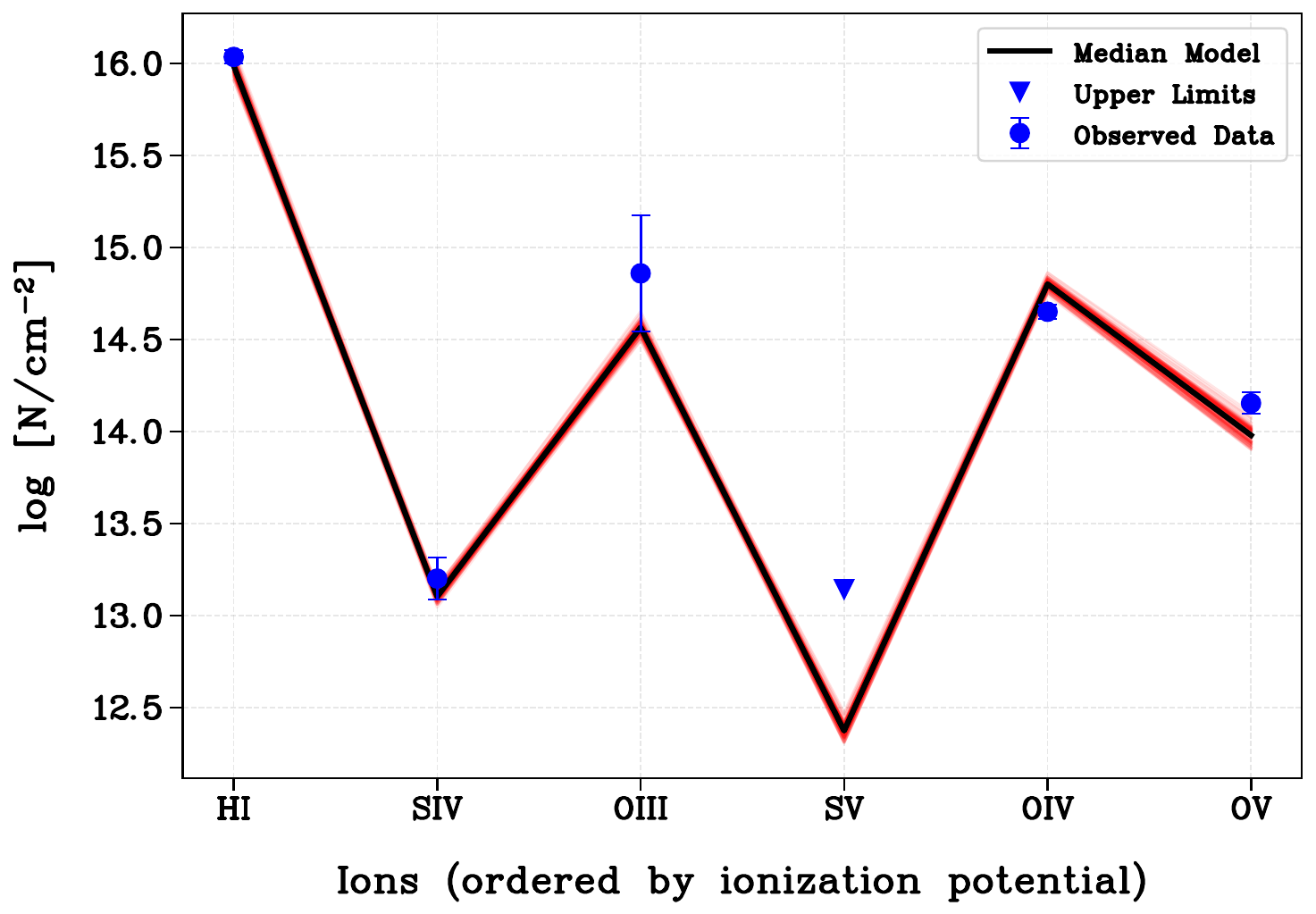}
\end{minipage}

\caption{\textbf{Left panel:} Posterior distributions from the MCMC \textsc{cloudy} modeling for the hydrogen number density, gas-phase metallicity, and \ion{H}{1} column density. The marginalized distributions yield median values with $1\sigma$ uncertainties:
$\log_{10}(n_{\mathrm{H}}/\mathrm{cm}^{-3})=-3.07\pm0.03$,
$\log_{10}(Z/Z_\odot)=-1.05\pm0.05$, and
$\log_{10}N(\text{\ion{H}{1}})=16.03\pm0.04$.
The two-dimensional posteriors show the 68\% and 95\% credible regions.
\textbf{Right panel:} Observed versus predicted ion column densities as a function of ionization potential. Blue points with $1\sigma$ uncertainties show the measured columns of \ion{H}{1}, \ion{S}{4}, \ion{O}{3}, \ion{O}{4}, and \ion{O}{5}, while the downward triangle denotes the upper limit for \ion{S}{5}. The black curve shows the best-fit \textsc{cloudy} model prediction and the red curves show realizations drawn from the posterior distribution. The model reproduces all constraining ions within their uncertainties.}
\label{fig:corner_ionisation}
\end{figure*}

\begin{figure}
\centering
\includegraphics[width=\linewidth]{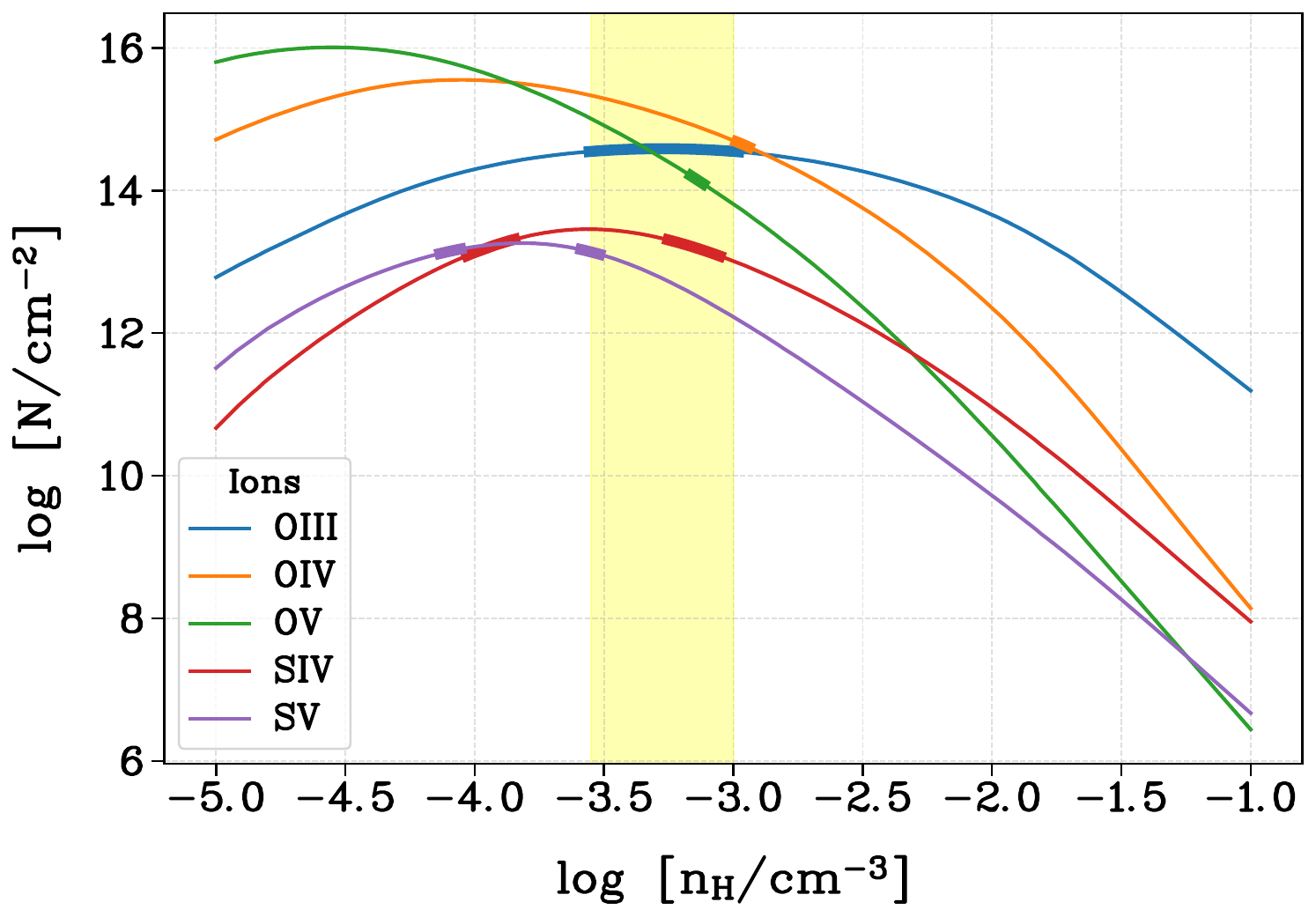}
\caption{Model-predicted column densities of \ion{O}{3}, \ion{O}{4}, \ion{O}{5} and \ion{S}{4} as a function of hydrogen number density at the best-fit metallicity of $\log_{10}(Z/Z_\odot) = -1.05$. The thickened segments on each curve mark the densities where the model column matches the observed column for that ion (within its 1$\sigma$ uncertainty). The intersection points for all ions are now consistent with a single hydrogen density of $\log_{10}( n_\text{H}/\mathrm{cm}^{-3}) \approx -3.07$, confirming that a single-phase model can simultaneously satisfy all observational constraints.}
\label{fig:col_density_vs_density}
\end{figure}

The MCMC analysis of our single-phase model yields a solution that reproduces the observed column densities of all constraining ions. The median values and $1\sigma$ confidence intervals from the posterior probability distributions are $\log_{10}(n_\text{H}/{\rm cm}^{-3}) = -3.07 \pm 0.03$, $\log_{10}(Z/Z_\odot) = -1.05 \pm 0.05$, and $\log_{10} N(\text{\ion{H}{1}}) = 16.03 \pm 0.04$. The posterior distributions are shown in Figure~\ref{fig:corner_ionisation}, and the derived physical parameters are listed in Table~\ref{tab:parameters_comp1}. Figure~\ref{fig:col_density_vs_density} illustrates the model-predicted column densities as a function of density. Despite the range of ionization potentials spanned by the detected ions, the single-phase model successfully recovers the observed columns within a narrow density range of $n_\text{H} \approx (0.3 - 1.0) \times 10^{-3}$~{\cc}, assuming a solar abundance pattern \citep{Grevesse}. We note that this derived density of $\log_{10}(n_\text{H}/\mathrm{cm}^{-3}) \approx -3.07$ suggests that this absorber traces relatively low-density gas in the outer CGM, consistent with its location at $\approx 0.9\,R_\mathrm{vir}$. The inferred gas density lies toward the lower end of the distribution reported for pLLSs and LLSs \citep[e.g.,][]{Lehner2019}, driven in part by the requirement to simultaneously reproduce the high-ionization \ion{O}{5} column density, which in a single-phase PIE framework demands a relatively high ionization parameter $U$, thereby driving the model toward low $n_\text{H}$. The simultaneous agreement of the oxygen and sulfur ions suggests that their relative abundance ratio (O/S) is consistent with the solar value, although the lack of other detected elements limits a broader assessment of the chemical pattern across all species. The model remains consistent with the \ion{S}{5} constraint, predicting a column density compatible with the observational upper limit. The significant agreement between the model predictions and observations suggests that the bulk of the absorbing gas can be adequately described by a single photoionized phase.

The non-detection of \ion{N}{4} $\lambda 765$ and the weak constraints from \ion{N}{3} provide critical insights into the chemical enrichment history of the absorber. Our best-fit photoionization model, assuming solar abundance ratios, predicts $\log_{10}[N(\text{\ion{N}{4}})] \approx 13.96$ and $\log_{10}[N(\text{\ion{N}{3}})] \approx 13.94$. However, the observed $3\sigma$ upper limit for \ion{N}{4} is $\log_{10}[N] \lesssim 12.96$, and the AOD measured upper limit for the contaminated \ion{N}{3} $\lambda 685$ feature is $\log_{10}[N] \lesssim 13.4$. This $\sim 1$~dex discrepancy indicates that the gas is under-enriched in nitrogen relative to oxygen ($[\text{N/O}] \lesssim -1.0$). Such sub-solar [N/O] ratios are a well-documented characteristic of low-metallicity environments ($\log_{10}(Z/Z_\odot) \lesssim -0.7$), where nitrogen production is dominated by primary nucleosynthesis in massive stars \citep{Pettini2002, Pettini2008}. The observed deficiency reflects the dual-timescale nature of nitrogen enrichment; while oxygen is produced by Type II supernovae, the secondary nitrogen component is released by intermediate-mass stars with a characteristic delay of $\gtrsim 250$~Myr \citep{Henry2000, Vangioni2018}. Consequently, the low [N/O] ratio in this pLLS suggests it traces gas in an early stage of chemical enrichment that has not yet been significantly enriched by the secondary nitrogen component. Finally, the model predicts $\log_{10}[N(\text{\ion{C}{2}})] \approx 13.1$, which is consistent with the observed upper limit value of $\log_{10} [N] \lesssim 14.1$ \citep[contaminated by Ly$\alpha$ at $z=0.060568$;][]{Tejos}, confirming that the model does not over-produce low-ionization carbon.

From Figure~\ref{fig:col_density_vs_density}, we can see that [O/H] cannot be lower than the 1/10th solar value predicted by the posterior distribution, since for lower metallicities the $N(\text{\ion{O}{3}})$ will not be reproduced for any density. The $N(\text{\ion{O}{4}})/N(\text{\ion{O}{5}})$ ratio is recovered for a density of $\log_{10}(n_\text{H}/{\rm cm}^{-3}) \approx -3.07$ given by the MCMC \textsc{cloudy} modeling. The model predicted $N(\text{\ion{O}{3}})$ is also consistent with the observed value at this density, as the \ion{O}{3} ionization fraction remains relatively insensitive to density in the range $-3.6 \lesssim \log_{10}(n_\text{H}/{\rm cm}^{-3}) \lesssim -2.7$.

Although the above solution provides a viable description of the absorber, it is important to note that the \ion{O}{3}, \ion{O}{4}, and \ion{O}{5}~ ions, whose ionization and creation potentials span a gradient, can also arise from a relatively narrow range of densities and photoionization temperatures, each contributing to the observed column densities of oxygen ions and hydrogen. Such an ionization-dependent stratification of the absorbing medium can be the reason for the narrower Doppler $b-$value for the \ion{O}{3}~ compared to \ion{O}{4}~ and \ion{O}{5}\footnote{The upper limit on the temperature set by the $b$ values of \ion{O}{5}, \ion{O}{4}, and \ion{O}{3}~ are $\text{T} \leq 8.7 \times 10^5$~K, $\text{T} \leq 4.2 \times 10^5$~K, and $\text{T} \leq 1.6 \times 10^5$~K respectively.}. Assuming this to be the case, the kinematic blending makes it difficult to partition the total column density of a given ion among the different density-temperature phases. 
The photoionization solution presented here should be regarded as an approximate representation of the physical conditions in the absorber, rather than a complete description of its ionization structure down to narrow (sub-kpc) physical or kinematic scales. Studies using synthetic spectra from simulated CGM sightlines have shown that such idealized single-phase models do recover the average metallicity and ionization conditions in the absorber, even when the real absorbing gas is more complex in its ionization structure at small scales \citep{Liang2018, Marra2021, Hafen2024}. 

\begin{table}
    \centering
    \renewcommand{\arraystretch}{1.2}
    \caption{Photoionization modeling Results}
    \label{tab:parameters_comp1}
    \begin{tabular}{lr}
        \hline
        \hline
        {Parameter} & {Comp} \\ \hline
        log$_{10}[n_\text{H}/\text{cm}^{-3}]$ & $-3.07^{+0.03}_{-0.03}$ \\ 
        log$_{10}[Z / Z_\odot]$ & $-1.05^{+0.05}_{-0.05}$ \\ 
        log$_{10}[\text{N(\ion{H}{1})}/\text{cm}^{-2}]$ & $16.0$ \\ 
        log$_{10}[\text{N(H)}/\text{cm}^{-2}]$ & $19.4$ \\ 
        $\text{P/k}$ (cm$^{-3}$K) & $40$ \\ 
        $\text{L}$ (kpc) & $9.7$ \\ 
        log$_{10}[\text{T/K}]$ & $4.3$ \\ \hline \hline
    \end{tabular}
    \renewcommand{\arraystretch}{1.0}
\end{table}

\section{GALAXIES ASSOCIATED WITH THE ABSORBER}
\label{sec:associated_galaxies}

\begin{figure*}
\centering

\includegraphics[width=0.95\textwidth]{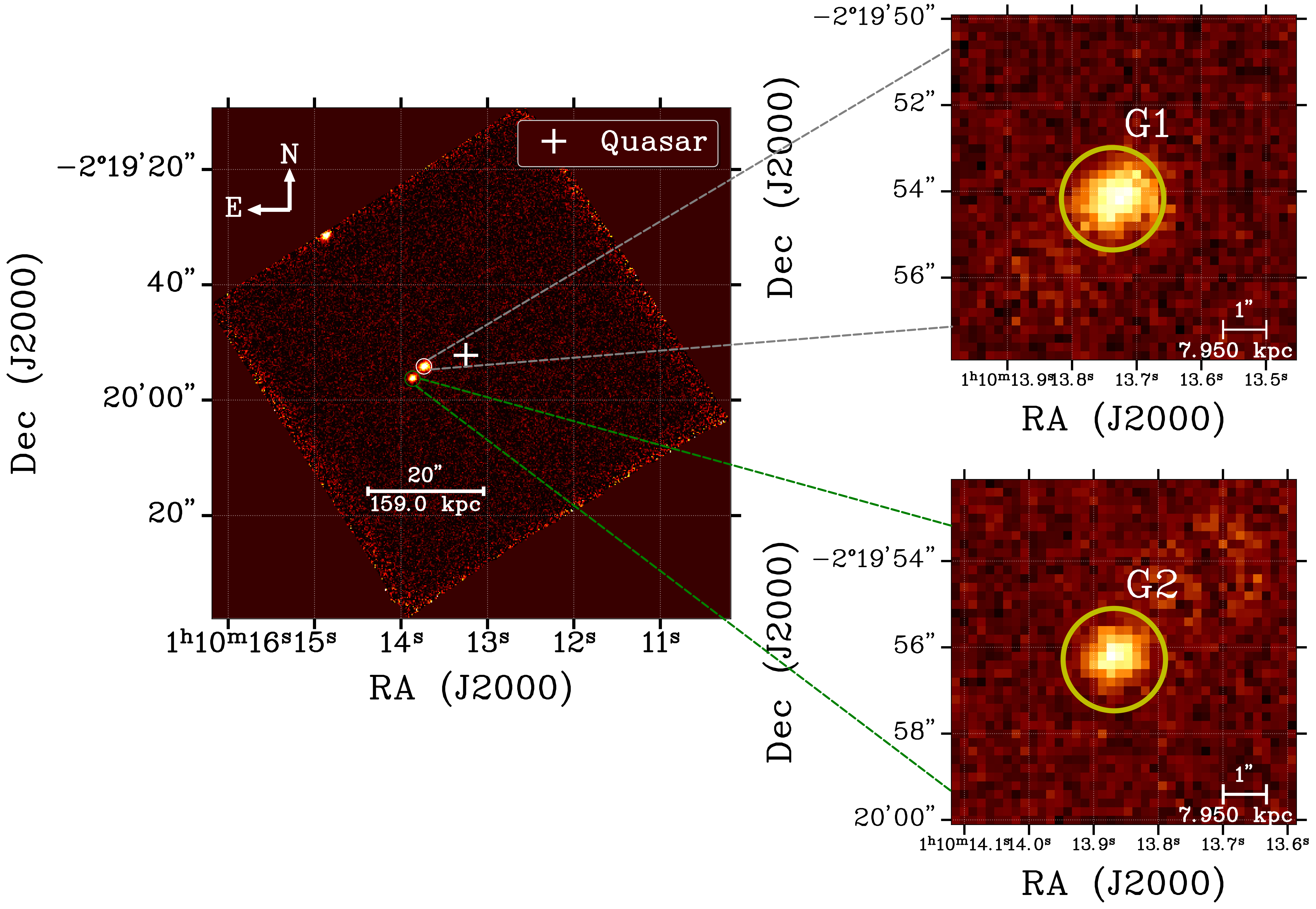}

\vspace{0.25cm}

\begin{minipage}[t]{0.48\textwidth}
    \centering
    \includegraphics[width=\linewidth]{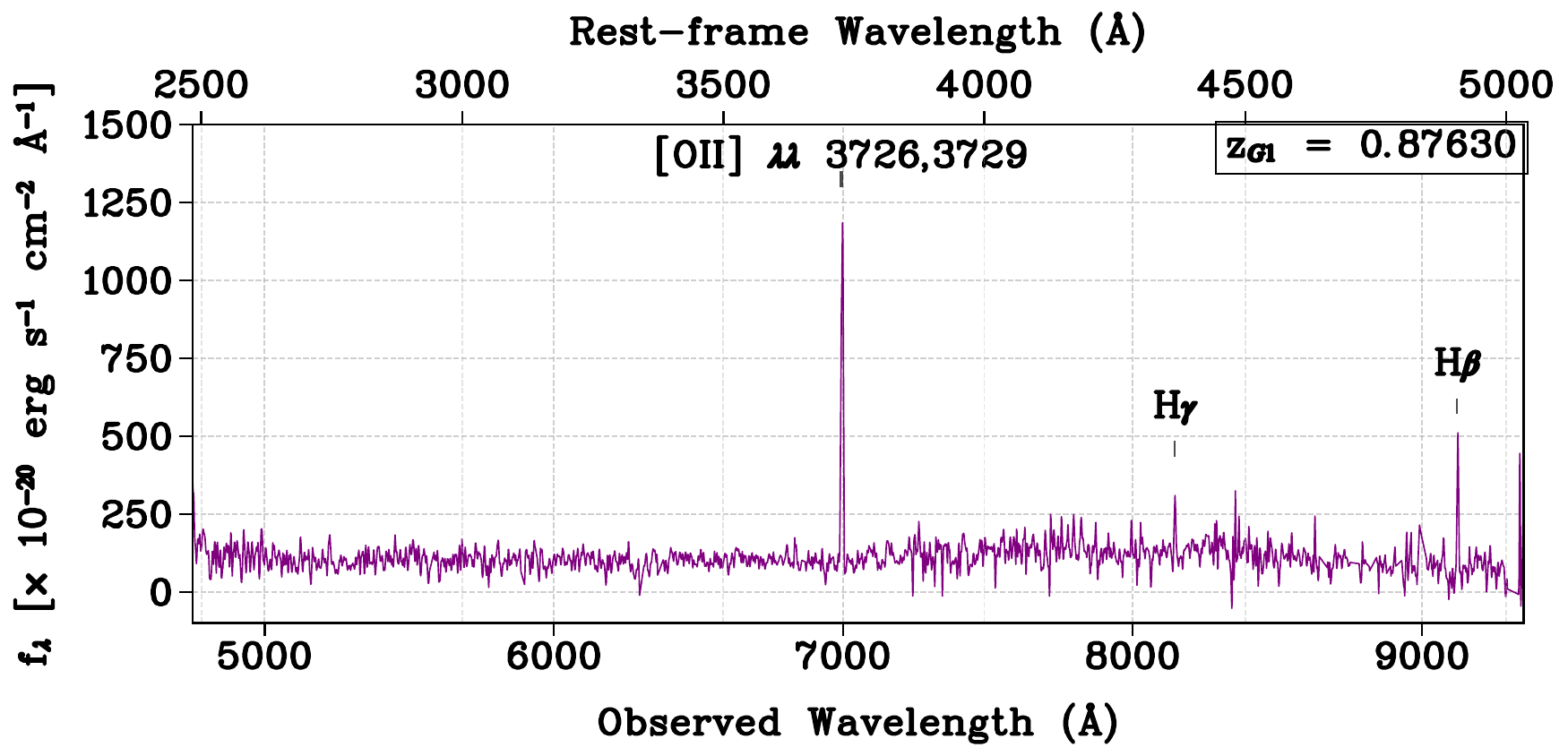}
\end{minipage}
\hfill
\begin{minipage}[t]{0.48\textwidth}
    \centering
    \includegraphics[width=\linewidth]{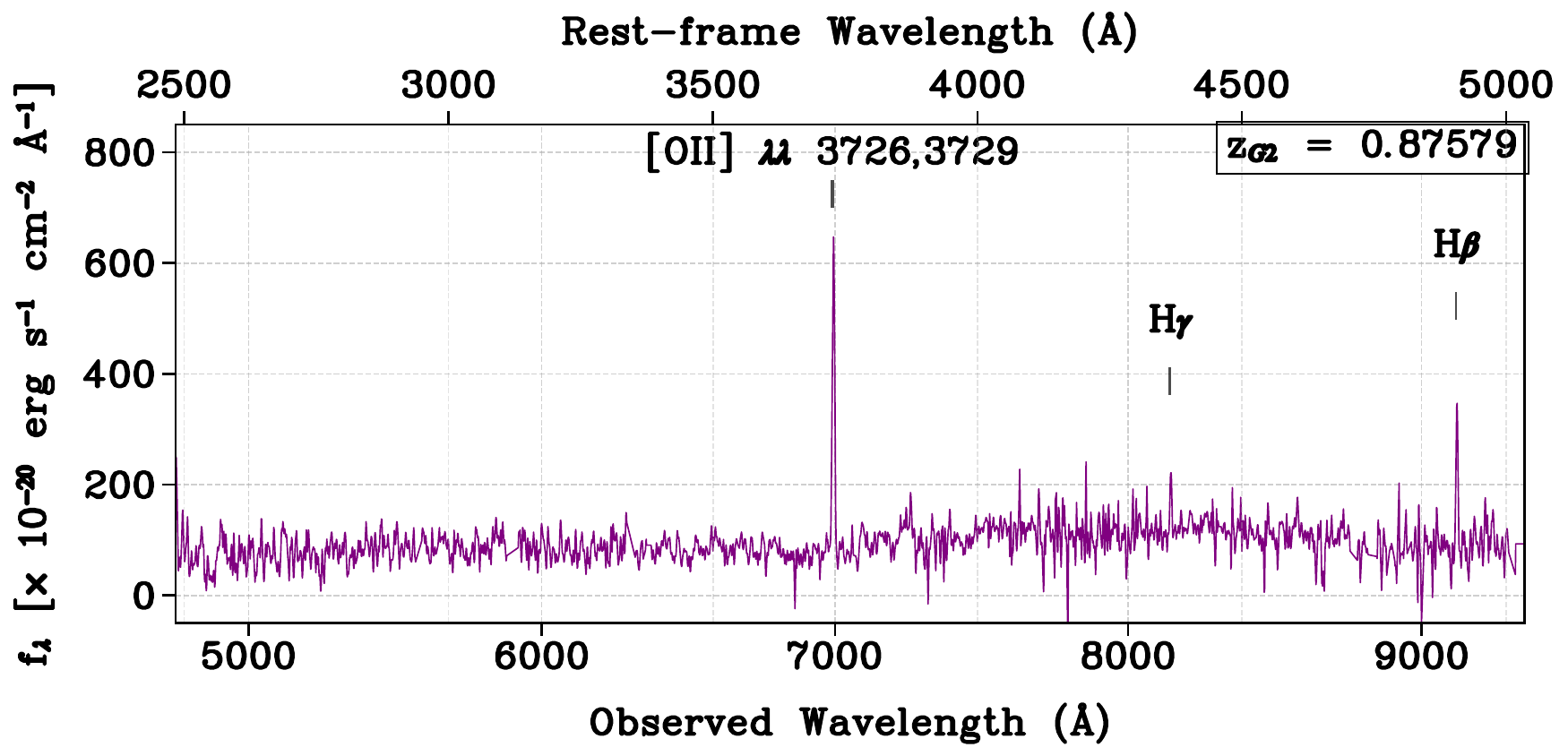}
\end{minipage}

\caption{Identification and characterization of the two galaxies, G1 and G2, associated with the absorption system at $z\approx0.876$. \textbf{Top:} Continuum- and quasar-PSF-subtracted narrowband [\ion{O}{2}] image from the full MUSE field of view, with zoomed galaxy insets. \textbf{Bottom:} Extracted one-dimensional spectra of Galaxy~1 (left) and Galaxy~2 (right), showing [\ion{O}{2}] $\lambda\lambda3726,3729$ and H$\beta$.}
\label{fig:combined_narrowband_spectra}
\end{figure*}

The raw MUSE spectra, originally in air wavelengths, were converted to vacuum wavelengths using the empirical formula from \citet{Filippenko_1982} to ensure kinematic consistency with the $HST$/COS data. As detailed in Section~\ref{sec:ifu_data}, we identified galaxy counterparts by searching the MUSE source catalog within a projected physical separation of $\leq 200$ pkpc and a line-of-sight velocity window of $\pm 250$~km~s$^{-1}$ relative to the absorber. This search identified two star-forming galaxies, G1 and G2, at redshifts of $z=0.87630$ and $z=0.87579$, respectively, placing them within the kinematic and spatial clustering scales typical of CGM associations. These galaxies lie at impact parameters of $\rho = 121$~kpc and $124$~kpc from the quasar sightline. Relative to the absorber redshift ($z=0.87641$), the line-of-sight velocity separations are $\Delta v = -18$ and $-99$~km~s$^{-1}$ for G1 and G2, respectively. The galaxies themselves have a physical separation of $22.6$~kpc (angular separation in the plane of the sky of $\approx 2.84^{\prime\prime}$). Both galaxy spectra show strong [\ion{O}{2}]~$\lambda\lambda$3726,3729 emission along with weaker H$\beta$ and H$\gamma$ lines, as shown in Figure~\ref{fig:combined_narrowband_spectra}.

\subsection{Morpho-kinematic Analysis}
\label{sec:galpak_analysis}
 
To model the galaxy kinematics and its orientation in the sky, we utilized the 3D line fitting package \galpak\ \citep{Bouche15}. Given the small angular separation between G1 and G2 ($\Delta \theta \approx 2.84^{\prime\prime}$), which can result in overlap of their light profiles, a mutual PSF subtraction was performed to isolate the light from each galaxy. Specifically, we subtracted the model light distribution of G1 from the sub-cube of G2, and vice-versa, to ensure the derived parameters were not biased by cross-contamination from the sources' extended wings. The algorithm fits a parametric disk model to the [\ion{O}{2}] emission, yielding constraints on the inclination of the galaxy with respect to the plane of the sky ($i$), the position angle of the galaxy major axis (PA), the maximum circular velocity of the emitting gas ($\text{v}_\text{max}$), the velocity dispersion of the emitting gas ($\sigma_\text{v}$), and the half-light radius ($R_{1/2}$). The \galpak results for galaxies G1 and G2 are listed in Table \ref{tab:galaxy_properties_all}, and their velocity maps and rotation curves are shown in Figure \ref{fig:combined_kinematics}

The \galpak\ modeling shows disk-like rotation for galaxies G1 and G2. From the best-fit kinematic parameters, we derived key physical properties for both galaxies, which are listed in Table~\ref{tab:galaxy_properties_all}. We derived the star-formation rates (SFR) using the empirical calibration from \citet{kewley_2004}. The intrinsic, reddening-corrected [\ion{O}{2}] luminosities were first calculated using their Equation (18), yielding L$_i = 4.16 \times 10^{42}~\text{erg s}^{-1}$ for G1 and L$_i = 1.77 \times 10^{42}~\text{erg s}^{-1}$ for G2. These luminosities were then used to determine the SFRs with Equation (4) from the same paper yielding ${\rm SFR}_{[\text{O~\textsc{ii}}]} = 27.4 \pm 6.9~\mathrm{M}_\odot\,\mathrm{yr}^{-1}$ for G1 and $11.7 \pm 3.0~\mathrm{M}_\odot\,\mathrm{yr}^{-1}$ for G2. The dynamical mass ($M_\text{dyn}$), halo mass ($M_\text{h}$), and stellar mass ($M_*$) were estimated using the scaling relations given in \citet{Bouche16, Girelli}. The stellar mass and SFR are additionally derived from SED modeling based on stellar population synthesis, as detailed in the next section.

\subsection{Stellar Population Synthesis}
\label{sec:prospector_analysis}

To constrain the stellar properties of G1 and G2, we performed a Bayesian Spectral Energy Distribution (SED) analysis on their extracted MUSE spectra using the \texttt{PROSPECTOR} code\footnote{Our choice of \texttt{PROSPECTOR} is motivated by its significant advantages over parametric codes. Its use of a flexible, non-parametric star-formation history (SFH) avoids the systematic biases on stellar mass and SFR that can be introduced by fixed functional forms. Furthermore, it is built upon the modern Flexible Stellar Population Synthesis (\textbf{FSPS}) framework and, crucially, allows the gas-phase metallicity to be fit independently of the stellar metallicity, providing a more physically realistic treatment of a galaxy's chemical enrichment history.} \citep{Johnson2021}. For our analysis, we adopted a non-parametric SFH with a continuity prior, the FSPS stellar population models with a \citet{Kroupa2001} initial mass function, and a two-component dust attenuation model. The resulting best-fit parameters and their credible intervals are summarized in Table~\ref{tab:galaxy_properties_all}. The stellar metallicity $\log_{10}(Z/Z_\odot)$ reported here is a mass-weighted quantity, as \texttt{PROSPECTOR} constructs the composite stellar population by integrating simple stellar populations weighted by the star-formation history in units of stellar mass \citep{Conroy2009, Johnson2021}. The spectral reconstruction and residuals for both galaxies are shown in Appendix Figures~\ref{fig:spectrum_residuals_G1} and \ref{fig:spectrum_residuals_G2}, and zoom-ins on the key emission lines are shown in Figures~\ref{fig:emission_lines_subplots_G1} and \ref{fig:emission_lines_subplots_G2}.

The \texttt{PROSPECTOR} fits constrain the stellar populations of both galaxies reasonably well, though constraining the gas-phase metallicity presents a well-understood challenge. The relatively large uncertainty in the gas-phase metallicity estimates arises from the limited emission-line coverage available in the $VLT$/MUSE spectra at this redshift. Only the \ion{O}{2} $\lambda\lambda3726,~3729$ doublet and H$\beta$ are detected, whereas the key metallicity-sensitive lines such as \ion{O}{3} $\lambda 5007$, H$\alpha$, and \ion{N}{2} $\lambda6584$ fall outside the observed wavelength coverage. Without these diagnostics, the well-known degeneracy in strong-line metallicity indicators cannot be fully resolved, leading to a significant covariance among metallicity, the ionization parameter, and the dust attenuation. Without these lines, the gas-phase metallicity cannot be tied to direct temperature-sensitive diagnostics, and the posteriors are correspondingly broad. In \texttt{PROSPECTOR}, the gas-phase metallicity is fit as a free parameter within the FSPS nebular emission framework \citep{Byler2017}, which self-consistently predicts emission line fluxes given the ionization parameter and gas metallicity. The posterior distributions for the gas-phase metallicity are therefore derived from the full MCMC sampling, and their width reflects the strong degeneracy between gas metallicity, ionization parameter, and dust attenuation in the absence of the key diagnostic lines noted above \citep{Curti2020}. No external strong-line calibration from \citet{Curti2020} was applied post-hoc; \citet{Curti2020} contextualizes the well-known degeneracy inherent to \ion{O}{2}$+$H$\beta$-only constraints, which the FSPS nebular framework of \citet{Byler2017} cannot fully resolve without additional line detections.

With this context, for galaxy G1, the analysis yields a near solar metallicity of $\log_{10}(Z/Z_\odot) = 0.10^{+0.04}_{-0.04}$ and gas-phase metallicity of $\log_{10}(Z_{\mathrm{gas}}/Z_\odot) = -0.04^{+0.28}_{-0.41}$. In contrast, Galaxy G2 shows a near-solar stellar metallicity of $\log_{10}(Z/Z_\odot) = 0.07^{+0.07}_{-0.06}$, and a lower median gas-phase metallicity of $\log_{10}(Z_{\mathrm{gas}}/Z_\odot) = -0.99^{+1.07}_{-1.33}$. The median values of the gas-phase metallicity for both galaxies are indicative of an ISM that is less enriched than the integrated stellar component, although the estimated uncertainties are large for the reasons mentioned earlier. Specifically, while the gas-phase metallicity of G1 is consistent with its near-solar stellar metallicity, the median value for G2 is approximately one-tenth of its stellar metallicity. Such a trend, if real, may reflect a variety of processes including ISM dilution by external gas, though the large uncertainties in the gas-phase metallicity make this interpretation tentative. We note, however, that the absorber metallicity of $\log_{10}(Z/Z_\odot) = -1.05$ is comparable to the median G2 gas-phase metallicity, so the absorber itself is not sufficiently metal-poor to drive further dilution of the ISM.

Cold-mode accretion models offer a natural framework for these properties. Cosmological simulations such as TNG50 show that for galaxies residing in halos below the stable virial shock threshold ($M_{\mathrm{h}} \lesssim 10^{12} \mathrm{M}_\odot$), the primary model of fueling is the accretion of cold, metal-poor gas \citep{Nelson2015, Nelson2019}. In this mass regime, gas flows along cosmic web filaments and is channeled directly onto the galactic disk along its major axis. This theoretical picture is corroborated by large observational campaigns like the \texttt{MEGAFLOW} survey, which finds a statistical link between major-axis \ion{Mg}{2} absorption--a tracer of inflowing gas--and galaxies undergoing enhanced star formation \citep{Zabl2019}. Taken together, the near-solar stellar metallicity of G2, the sub-solar median gas-phase metallicity broadly consistent with that of the absorber, and the absorber's alignment with the major axis represent a self-consistent set of properties that cold-mode accretion can naturally account for. However, given that G1 and G2 are separated by only $22.6$~kpc in projection -- placing them well within each other's virial radii -- a dynamical origin in which the absorbing gas traces tidally displaced or intragroup material arising from the ongoing interaction of the pair remains a physically motivated alternative that the present data cannot exclude.

To place these properties in their evolutionary context, we compare the galaxies to the star-forming main sequence (SFMS). We use the comprehensive parameterization from \citet{Speagle2014}, which defines the relation as:
\[
\log_{10}(\mathrm{SFR}) = (0.84 - 0.026\,t)\times \log_{10}(M_*)-(6.51 - 0.11\,t)
\]
where $t$ is the age of the Universe in Gyr. For our adopted cosmology at the galaxies' redshift of $z \approx 0.876$, the age of the Universe is $t \approx 6.39\,\mathrm{Gyr}$. Using the \texttt{PROSPECTOR} stellar masses, the \citet{Speagle2014} parameterization predicts main-sequence SFRs of $\approx 6.3\,\mathrm{M}_\odot\,\mathrm{yr}^{-1}$ (G1) and $\approx 14.5\,\mathrm{M}_\odot\,\mathrm{yr}^{-1}$ (G2). Comparing the instantaneous star formation rates derived from the [\ion{O}{2}] emission line \citep{kewley_2004} to the predicted main-sequence values from \citet{Speagle2014}, we find that G1 exhibits an $\mathrm{SFR}_{[\text{O~\textsc{ii}}]} = 27.4 \pm 6.9\,\mathrm{M}_\odot\,\mathrm{yr}^{-1}$. This corresponds to an offset of $\Delta\log_{10}(\mathrm{SFR}) \simeq +0.64\,\mathrm{dex}$ above the MS ridge, placing it $\approx 2.1\sigma$ above the relation (assuming a $0.3\,\mathrm{dex}$ scatter) and characterizing it as an actively star-forming system at this epoch. For G2, the $\mathrm{SFR}_{[\text{O~\textsc{ii}}]} = 11.67 \pm 2.95\,\mathrm{M}_\odot\,\mathrm{yr}^{-1}$ yields $\Delta\log_{10}(\mathrm{SFR}) \simeq -0.09\,\mathrm{dex}$, placing it well within the $1\sigma$ intrinsic scatter of the MS. The notable discrepancy between these instantaneous rates and the 100-Myr-averaged \texttt{PROSPECTOR} values ($2.5$ and $2.2\,\mathrm{M}_\odot\,\mathrm{yr}^{-1}$) is a consequence of the different timescales and physical tracers involved. While [\ion{O}{2}] emission is sensitive to the ionizing radiation from massive stars with lifetimes $\lesssim 10\,\mathrm{Myr}$, the SED fitting in \texttt{PROSPECTOR} integrates the star formation history over a longer interval, smoothing out short-term fluctuations in star formation. The [\ion{O}{2}] tracer is highly dependent on the assumed nebular extinction and ionization state, whereas \texttt{PROSPECTOR} constrains the SFR through the global stellar continuum and a flexible dust attenuation law \citep{kewley_2004, Speagle2014, Johnson2021}.

With the instantaneous star formation rates placing G1 in a regime of elevated star formation and G2 firmly on the star-forming main sequence, both galaxies are confirmed as highly active systems. The presence of strong [\ion{O}{2}] and H$\beta$ emission lines is entirely consistent with these findings. While the 100-Myr-averaged rates from \texttt{PROSPECTOR} are lower, the current gas-phase activity (as traced by [\ion{O}{2}]) is the more relevant metric for assessing the galaxies' immediate impact on their circumgalactic environment \citep{kewley_2004, Speagle2014}.

Finally, we compare the stellar masses derived from SED fitting with those estimated from kinematic scaling relations to assess systematic uncertainties. A notable discrepancy is observed: for G1, the stellar mass inferred from the $M_* - V_{\text{max}}$ relation \citep{Bouche16} is higher by a factor of $\sim4.4$, and for G2, it is higher by a factor of $\sim1.4$. This difference arises from the distinct physical properties and assumptions probed by each technique. SED fitting is fundamentally a measure of the stellar mass-to-light ratio, making it highly sensitive to the assumed star-formation history, dust content, and initial mass function. In contrast, kinematic scaling relations are calibrated to statistically isolate the stellar component from the total dynamical mass enclosed within a characteristic radius, making them sensitive to the assumed dark matter fraction and the intrinsic scatter in the empirical relations themselves. While the exact values differ, both methods yield stellar masses of the same order of magnitude, $\sim 10^{10}\,\mathrm{M}_\odot$. At this epoch, this mass scale places them near the characteristic stellar mass of the star-forming galaxy population \citep{Whitaker2014}. It is galaxies within this mass range that are predicted and observed to host the most extensive and dynamically active circumgalactic media \citep{Tumlinson2017}. Therefore, despite the systematic uncertainties in determining their precise stellar mass, the convergence of both methods on this critical mass scale strongly supports the combined halos of the G1-G2 galaxy pair as the potential host of the absorber.

\subsection{Are the Galaxies Driving Winds?}
\label{sec:outflow_capability}

As both galaxies exhibit ongoing star formation, we assess whether their current activity is sufficient to drive large-scale galactic winds capable of enriching CGM baryonic content through the recycling of metal-enriched gas. The ability of galaxies to drive such outflows is closely tied to their star formation surface density ($\Sigma_{\text{SFR}}$), defined as: 
\begin{equation*}
    \Sigma_{\text{SFR}} = \frac{0.5 \times \text{SFR}}{\pi R_{1/2}^2} = \frac{\text{SFR}}{2\pi R_{1/2}^2}
\end{equation*}
\noindent where SFR is the star formation rate and $R_{1/2}$ is the half-light radius. Using the instantaneous star formation rates derived from [\ion{O}{2}] ($27.37\,\mathrm{M}_\odot\,\mathrm{yr}^{-1}$ for G1 and $11.67\,\mathrm{M}_\odot\,\mathrm{yr}^{-1}$ for G2) and the radii from Table~\ref{tab:galaxy_properties_all}, we calculate $\Sigma_{\text{SFR}} \approx 0.098\,\mathrm{M}_\odot\,\mathrm{yr}^{-1}\,\mathrm{kpc}^{-2}$ for G1 and $\Sigma_{\text{SFR}} \approx 0.071\,\mathrm{M}_\odot\,\mathrm{yr}^{-1}\,\mathrm{kpc}^{-2}$ for G2. 

G1 approaches the classical threshold of $\Sigma_{\text{SFR}} \gtrsim 0.1\, \mathrm{M}_\odot\,\mathrm{yr}^{-1}\,\mathrm{kpc}^{-2}$ established for driving powerful, large-scale outflows in local starbursts \citep{Heckman2002}. G2 also significantly exceeds the critical threshold of $\Sigma_{\text{SFR}} \sim 0.01\,\mathrm{M}_\odot\,\mathrm{yr}^{-1}\,\mathrm{kpc}^{-2}$ suggested for $z \sim 2$ galaxies to drive gas out of their gravitational potentials \citep{Martin2012, RobertsBorsani2020}. These results indicate that both galaxies are, in principle, capable of driving metal-enriched winds. However, the absorber metallicity of $\log_{10}(Z/Z_\odot) = -1.05$ -- while sub-solar and well below the stellar metallicities of either galaxy -- lies above the median for $z < 1$ pLLSs of $\log_{10}(Z/Z_\odot) = -1.3$ \citep{Lehner2019, Wotta2019}, where the metallicity distribution spans the range $-3.0 \lesssim \log_{10}(Z/Z_\odot) \lesssim 0.4$, and is therefore not characteristic of near-pristine IGM accretion. Rather, it occupies the intermediate metallicity regime where accreting filamentary gas, moderately pre-enriched by prior episodes of star formation, is indistinguishable in metallicity from gas tidally stripped or dispersed through the dynamical interaction of a close galaxy pair. The major-axis alignment is consistent with an inflow geometry \citep{Ho2016, Prusinski2021}, but is not sufficient on its own to uniquely identify the gas as a cold accretion stream.

\begin{table*}
\centering
\caption{Summary of properties for G1 and G2. For each galaxy, the top section lists properties derived from \galpak\ morpho-kinematic modeling and associated scaling relations; direct \galpak\ outputs are marked with a star ($\star$). The bottom section lists stellar population properties derived from stellar population synthesis fitting with \texttt{PROSPECTOR}, marked with a dagger ($\dagger$).}
\label{tab:galaxy_properties_all}

\renewcommand{\arraystretch}{1.5}

\begin{minipage}[t]{0.49\textwidth}
\centering
\begin{tabular}{l @{} >{\hspace{10pt}}c}
\hline\hline
Property & Value (G1) \\
\hline
z & $0.87630~\pm~0.00003$ \\
    RA (J2000) & 01:10:13.73 \\
    Dec (J2000) & $-$02:19:54.17 \\
    Flux$_{\text{([O~\textsc{ii}])}}$ (10$^{-17}$ erg s$^{-1}$ cm$^{-2}$) $^\star$ & $9.03^{+0.15}_{-0.15}$ \\
    $\text{R}_{1/2}$ (kpc) $^\star$ & $6.65^{+0.13}_{-0.13}$ \\
    $\text{R}_{\text{turnover}}$ (kpc) $^\star$ & $3.55^{+0.75}_{-0.75}$ \\
    Inclination ($^{\circ}$) $^\star$ & $35.1^{+1.5}_{-1.5}$ \\
    P.A. ($^{\circ}$) $^\star$ & $110.6^{+2.3}_{-2.3}$ \\
    $\text{V}_{\text{max}}$ (km s$^{-1}$) $^\star$ & $152.23^{+2.66}_{-2.66}$ \\
    Velocity Dispersion V$_\text{disp}$ (km s$^{-1}$) $^\star$ & $41.54^{+1.49}_{-1.49}$ \\
    Escape Velocity $\text{V}_{\text{escape}}$ (km s$^{-1}$) & $215.52^{+4.71}_{-4.71}$ \\
    $\text{SFR}~(\text{M}_\odot\,\text{yr}^{-1})$ & $27.37^{+6.90}_{-6.90}$ \\
    M$_\text{dyn}$ ($10^{10}~\text{M}_{\odot}$) & $3.57^{+0.14}_{-0.14}$ \\
    M$_\text{h}$ ($10^{11}~\text{M}_\odot$) & $7.24^{+0.38}_{-0.38}$ \\
    M$_*$ ($10^{10}~\text{M}_\odot$) & $2.77^{+0.30}_{-0.30}$ \\
    Azimuthal angle $(\alpha)(^{\circ})$ & $9.00^{+2.3}_{-2.3}$ \\
    R$_\text{vir}$ (kpc) & $135.72^{+2.86}_{-2.86}$\\
    $\rho/$R$_\text{vir}$ & $0.888$ \\
    R$_{200}$ (kpc) & $135.40^{+2.59}_{-2.59}$ \\
    \hline
\multicolumn{2}{c}{\textit{Stellar Population Properties} \texttt{(PROSPECTOR)}} \\
    \hline
    $\mathrm{SFR}_{100\,\mathrm{Myr}}$ (M$_\odot\,\mathrm{yr}^{-1}$)~$^{\dagger}$ & $2.46^{+0.69}_{-0.46}$ \\
    M$_*~(10^9~\text{M}_\odot)$~$^{\dagger}$ & $6.31^{+1.56}_{-1.46}$ \\
    $\text{log}_{10}(Z_{\mathrm{gas}}/Z_\odot)$~$^{\dagger}$ & $-0.04^{+0.28}_{-0.41}$ \\
    $\text{log}_{10}(Z/Z_\odot)$~$^{\dagger}$ & $0.10^{+0.04}_{-0.04}$ \\
    \hline
\end{tabular}
\end{minipage}
\hfill
\begin{minipage}[t]{0.49\textwidth}
\centering
\begin{tabular}{l @{} >{\hspace{10pt}}c}
\hline\hline
Property & Value (G2) \\
\hline
z & $0.87579~\pm~0.00005$ \\
    RA (J2000) & 01:10:13.86 \\
    Dec (J2000) & $-$02:19:56.16 \\
    Flux$_{\text{([O~\textsc{ii}])}}$ ($10^{-17}$ erg s$^{-1}$ cm$^{-2}$) $^\star$ & $5.11^{+0.12}_{-0.12}$ \\
    $\text{R}_{1/2}$ (kpc) $^\star$ & $5.12^{+0.16}_{-0.16}$ \\
    $\text{R}_{\text{turnover}}$ (kpc) $^\star$ & $1.43^{+0.56}_{-0.56}$ \\
    Inclination ($^{\circ}$) $^\star$ & $28.7^{+2.8}_{-2.8}$ \\
    P.A. ($^{\circ}$) $^\star$ & $115.0^{+4.0}_{-4.0}$ \\
    $\text{V}_{\text{max}}$ (km s$^{-1}$) $^\star$ & $156.36^{+8.30}_{-8.30}$ \\
    Velocity Dispersion V$_\text{disp}$ (km s$^{-1}$) $^\star$ & $66.91^{+4.50}_{-4.50}$ \\
    Escape Velocity $\text{V}_{\text{escape}}$ (km s$^{-1}$) & $221.54^{+12.67}_{-12.67}$ \\
    $\text{SFR}~(\text{M}_\odot\,\text{yr}^{-1})$ & $11.67^{+2.95}_{-2.95}$ \\
    M$_\text{dyn}$ ($10^{10}~\text{M}_{\odot}$) & $2.91^{+0.32}_{-0.32}$ \\
    M$_\text{h}$ ($10^{11}~\text{M}_\odot$) & $7.85^{+1.25}_{-1.25}$ \\
    M$_*$ ($10^{10}~\text{M}_\odot$) & $3.09^{+0.71}_{-0.71}$ \\
    Azimuthal angle $(\alpha)(^{\circ})$ & $5.2^{+4.0}_{-4.0}$ \\
    R$_\text{vir}$ (kpc) & $139.40^{+7.58}_{-7.58}$\\
    $\rho/$R$_\text{vir}$ & $0.887$ \\
    R$_{200}$ (kpc) & $138.97^{+7.46}_{-7.46}$ \\
    \hline
\multicolumn{2}{c}{\textit{Stellar Population Properties} \texttt{(PROSPECTOR)}} \\
\hline
    $\mathrm{SFR}_{100\,\mathrm{Myr}}$ (M$_\odot\,\mathrm{yr}^{-1}$)~$^{\dagger}$ & $2.23^{+0.86}_{-0.47}$ \\
    M$_*~(10^{10}~\text{M}_\odot)$~$^{\dagger}$ & $2.19^{+0.50}_{-0.53}$ \\
    $\text{log}_{10}(Z_{\mathrm{gas}}/Z_\odot)$~$^{\dagger}$ & $-0.99^{+1.07}_{-1.33}$ \\
    $\text{log}_{10}(Z/Z_\odot)$~$^{\dagger}$ & $0.07^{+0.07}_{-0.06}$ \\
    \hline
\end{tabular}
\end{minipage}

\renewcommand{\arraystretch}{1.0}
\end{table*}

\begin{figure*}
\centering

\begin{minipage}[t]{0.48\textwidth}
    \centering
    \includegraphics[width=\linewidth]{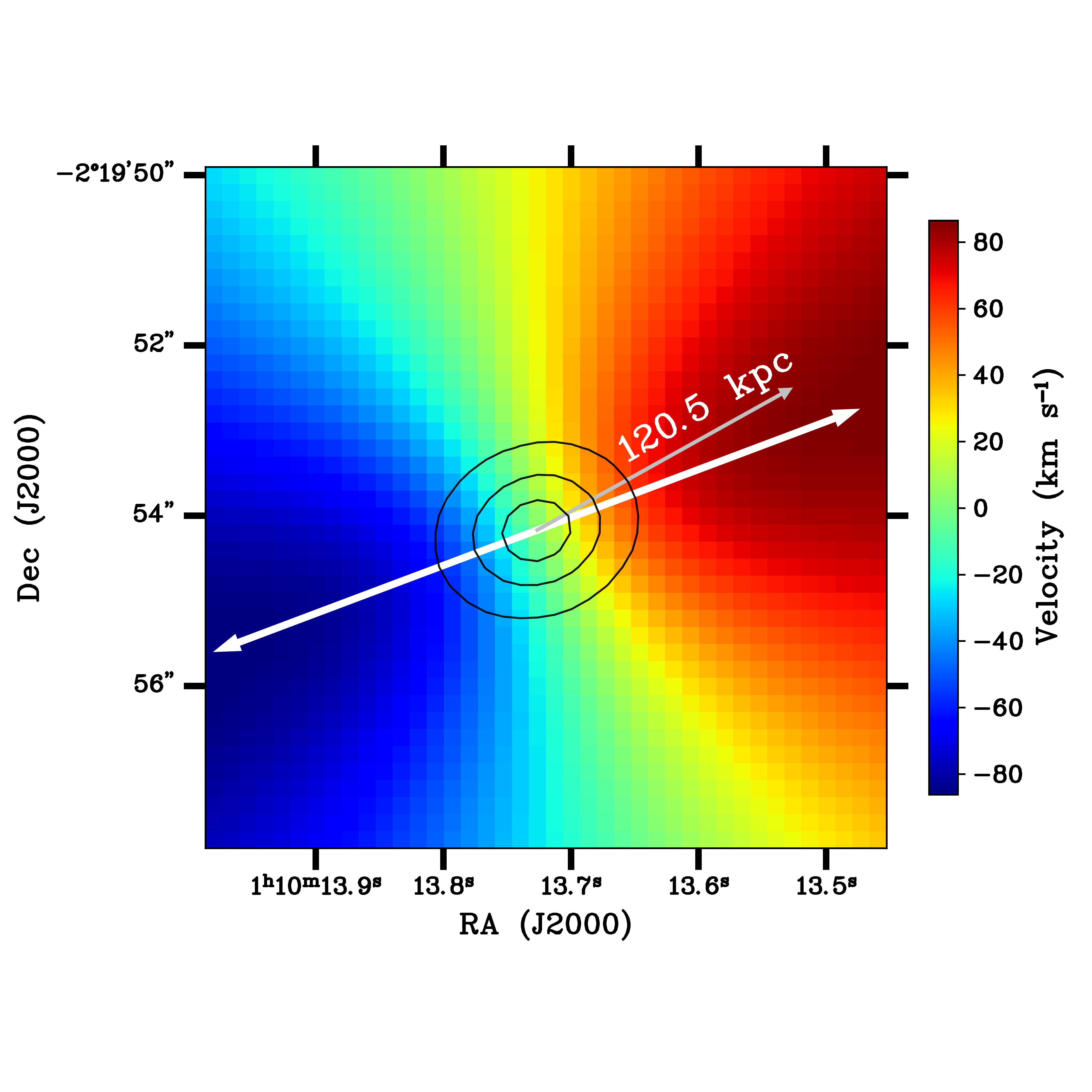}
\end{minipage}
\hfill
\begin{minipage}[t]{0.48\textwidth}
    \centering
    \includegraphics[width=\linewidth]{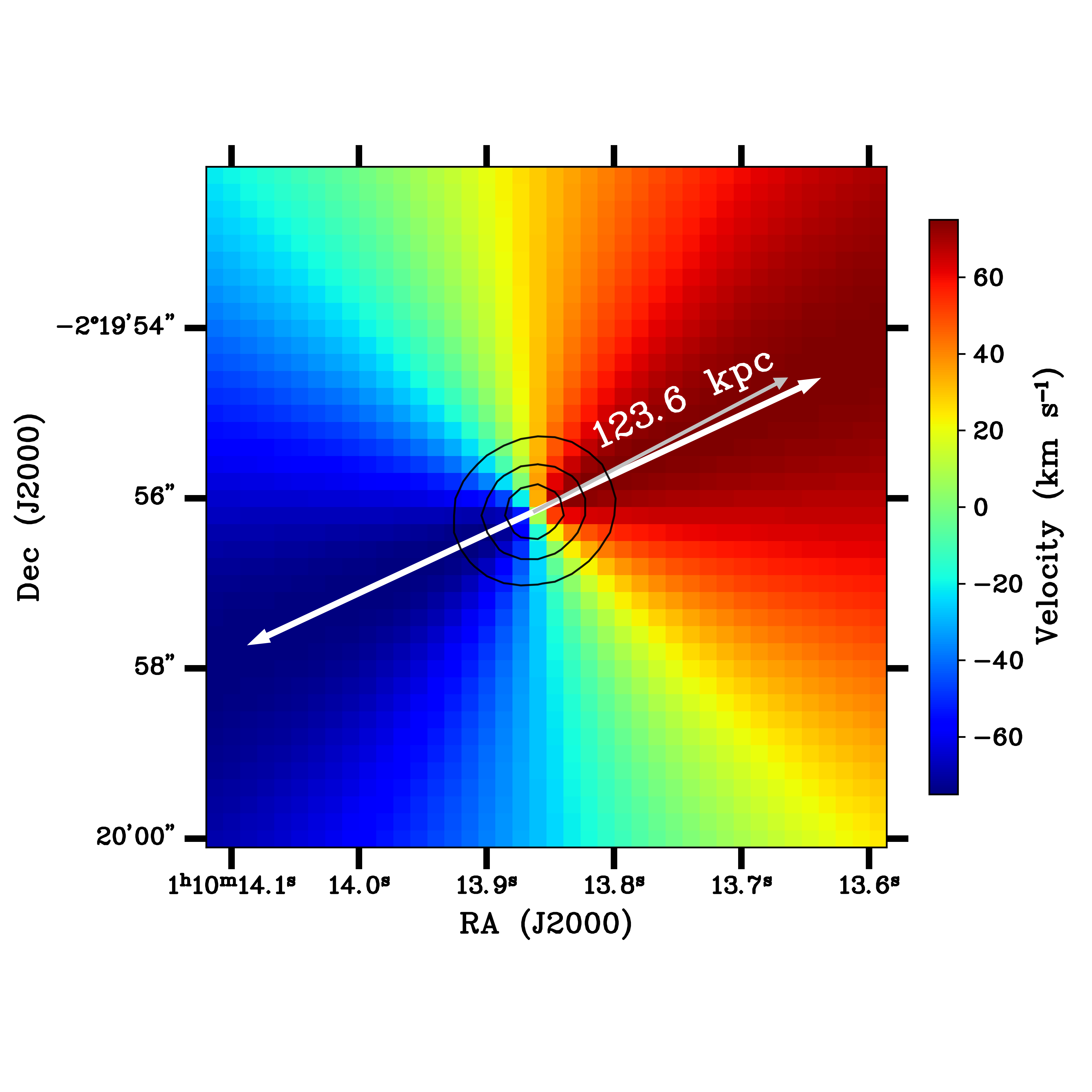}
\end{minipage}

\vspace{0.25cm}

\begin{minipage}[t]{0.48\textwidth}
    \centering
    \includegraphics[width=\linewidth]{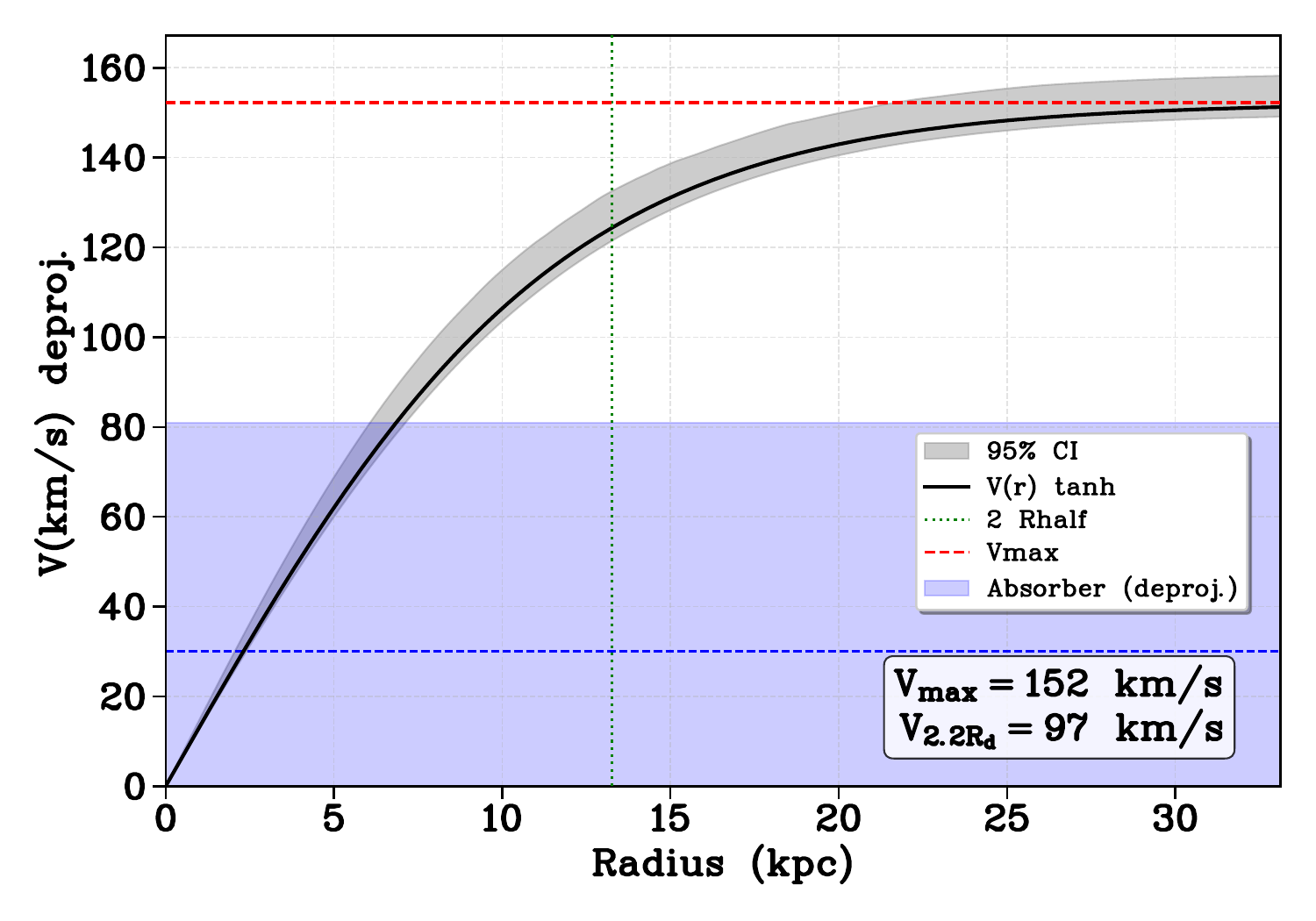}
\end{minipage}
\hfill
\begin{minipage}[t]{0.48\textwidth}
    \centering
    \includegraphics[width=\linewidth]{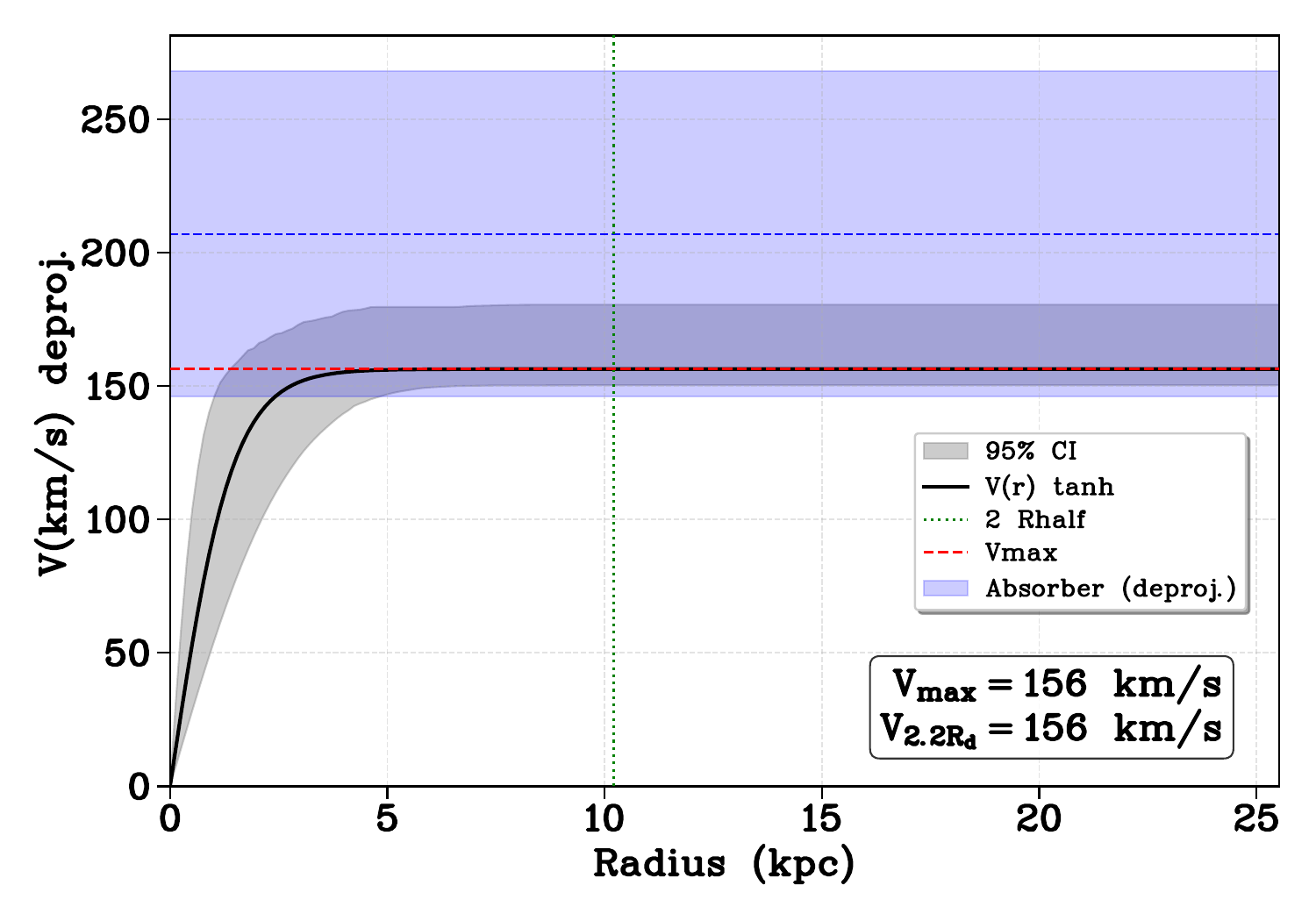}
\end{minipage}

\caption{Morpho-kinematic properties of Galaxy~1 (left column) and Galaxy~2 (right column).
\textbf{Top row:} Velocity maps derived from \galpak\ modeling of the [\ion{O}{2}] emission. The white line marks the major axis, the \textcolor{mygray}{gray} line indicates the direction to the quasar sightline, and ellipsoids show iso-flux contours. The projected distances are 120.5~kpc (G1) and 123.6~kpc (G2).
\textbf{Bottom row:} Deprojected rotation curves from \textsc{GaLPaK}$^{\mathrm{3D}}$. The rotation curves are constructed by \galpak\ fitting a parametric tanh disk model to the full three-dimensional emission-line data cube, which simultaneously fits the spatial and spectral dimensions to recover the intrinsic, inclination-corrected circular velocity as a function of radius. The black solid line shows the best-fit tanh model; the \textcolor{mygray}{gray} band indicates the 95\% confidence interval. The red dashed line marks $V_{\mathrm{max}}$, and the vertical green dotted line shows $2R_{\mathrm{half}}$. The blue dashed line and shaded band show the absorber velocity and $b$ parameter. The QSO sightline lies at projected distances of 120.5~kpc (G1) and 123.6~kpc (G2), corresponding to $\approx 0.9\,R_\mathrm{vir}$, well beyond the x-axis range shown. At this radius, an NFW halo profile predicts a circular velocity declining well below $V_\mathrm{max}$, with the vast majority of the halo mass already enclosed within $0.9\,R_\mathrm{vir}$.}
\label{fig:combined_kinematics}
\end{figure*}

\section{ORIGIN OF THE ABSORBER}
\label{sec:origin_of_absorption}

The small azimuthal angles ($\alpha \approx 5^{\circ}-9^{\circ}$) place the absorber along the projected major axis of both galaxies. Studies of \ion{Mg}{2} absorbers have established a bimodal azimuthal angular dependence, characterized by an excess of absorption detected along the major and minor axes of host galaxies \citep{Bouche12, Kacprzak_2012, Zabl2019}. Enhanced absorption along the projected major axis is typically attributed to the line of sight intersecting extended, rotating gaseous structures or accreting material \citep{Bordoloi2011, Ho17, Zabl2019}, whereas absorption detected along the minor axis is associated with large-scale, biconical outflows driven by stellar feedback processes \citep{Veilleux_2005, Schroetter2019}. While early surveys of Lyman limit systems suggested a distinct metallicity bimodality in the circumgalactic medium \citep{Lehner2013, Wotta_2016}, more recent analyses based on larger samples suggest a more continuous or unimodal distribution, noting that gas-phase metallicity may not depend strongly on azimuthal orientation \citep{Pointon2019, Sameer2024}. Regardless of the global metallicity distribution, the close alignment of the current absorber with the projected major axes of G1 and G2 suggests a physical connection with inflowing or recycled material accreting onto the galaxy pair.

Given this geometrical orientation, we examine whether the absorber shares kinematics consistent with disk rotation. As seen from the rotation curves in Figure~\ref{fig:combined_kinematics}, the deprojected velocity of the absorber is $\approx 30$~km~s$^{-1}$ for G1 and $\approx 207$~km~s$^{-1}$ for G2. While the absorber's line-of-sight motion follows the same rotational sense as the disks of both galaxies, the deprojected velocity offsets -- derived from the measured line-of-sight separations of $|\Delta v| \approx 18$ and $99$~km~s$^{-1}$ -- are inconsistent with a straightforward continuation of the disk rotation. These values show a significant departure from the rotation plateaus of the respective galaxies ($V_{\text{max}} \approx 152$ and $156$~km~s$^{-1}$). The absorber lies at a projected separation of $\sim 0.9~R_\text{vir}$ from either galaxy, placing it in the outer halo where the kinematic coupling between the halo gas and the central disk rotation is expected to be weak. As simulations show, beyond $\approx 0.3~R_\text{vir}$, the motion of the halo gas becomes uncorrelated with that of the central disk, exhibiting a greater degree of kinematic scatter \citep{Stewart2013, Nelson2015}.

Adopting an average halo mass of $M_\text{h} = 7.5 \times 10^{11}$~M$_\odot$, and virial radius of R$_\text{vir} = 137$ kpc\footnote{Galaxies G1 and G2 have similar halo masses and virial radii. Hence, we use the average of the two values.}, we estimate the virial temperature at 0.9R$_\text{vir}$ to be $T_\text{vir} \approx 8.5 \times 10^5$~K. This temperature exceeds the value inferred from photoionization modeling by nearly a factor of forty, indicating that the absorbing gas is not in thermal equilibrium with the virialized halo medium. Instead, it likely represents cooler material not shock-heated to the virial temperature. Such a scenario is consistent with cold-mode accretion, which is predicted to dominate the mechanism by which galaxies acquire gas in halos with masses below the critical threshold of $M_\text{h} \approx 5 \times 10^{12}$~M$_\odot$ needed to sustain a stable virial shock \citep{Keres2005, Keres2009, Dekel2006, Dekel2009, Faucher_Giguère_2011}, though we note that sub-virial temperatures are also expected for tidally stripped or intragroup gas in close galaxy-pair environments, or for gas condensing out of a cooling hot halo atmosphere \citep{Maller2004, Keres2009}. 

Cold-mode accretion is thought to occur through intergalactic filaments or clumps that penetrate the halos of galaxies. Once accreted, simulations suggest that the infalling gas rapidly reaches pressure equilibrium through compression by the surrounding hot halo medium, while remaining relatively cooler and denser than the ambient gas \citep{vandeVoort2012}. To examine whether the absorber could be in pressure confinement by the hot halo, we make an estimate based on the parametric density profile (the spherical $\beta$ model) for the hot halos of galaxies proposed by \citet{Miller2013, Miller2015}. Using this prescription, we find the number density of hot plasma at the virial radius to be $n(R_{\mathrm{vir}}) \approx (4.6-23.2) \times 10^{-5}$~cm$^{-3}$, corresponding to a thermal pressure of $P/k \approx (39-197)$~cm$^{-3}$ K (the details of this calculation are included in the Appendix \ref{sec:appendixA}). These values are comparable to the estimated pressure of the absorbing gas, $P/k \approx 40$~cm$^{-3}$ K, derived from the photoionization model, indicating that the absorbing gas is in near pressure equilibrium with the $T \approx 10^6$~K halo expected to be present around either galaxy. The range in halo pressure is based on the plausible values for the hot-gas fraction, $f_{\mathrm{hot}} \approx 0.1$–$0.5$, for halos with masses of $M_{\mathrm{h}} \sim 10^{11}$–$10^{12}$~M$_\odot$ \citep{Stern2016, Pandya2021}. For lower mass halos, $f_\mathrm{hot}$ will be less. This fraction directly influences the normalization of the model density profile. The slope parameter $\beta$ in the model, which governs the radial decline of the gas density 
can vary across systems. In the calculation, we adopt a representative value of $\beta = 0.5$. The resulting pressure at the virial radius should therefore be regarded as an order-of-magnitude estimate rather than a precise measurement.

Another key diagnostic for understanding the origin of the absorber is its metallicity relative to that of the associated galaxies. Photoionization modeling constrains the absorber metallicity to approximately one-tenth solar. Although the gas-phase metallicity estimates for both galaxies carry significant uncertainties (see Sec. \ref{sec:prospector_analysis}), the median ISM metallicity for G1 is fully consistent with its near-solar stellar metallicity within the $1\sigma$ errors. The median of the posterior distribution suggests that the ISM metallicity of G2 may be lower than that of its stellar component by a factor of approximately 11. However, the associated uncertainties are large, and the two measurements are statistically consistent within errors. If real, such a difference could be indicative of recent dilution of the ISM by metal-poor gas \citep{FaucherGiguereKeresMa2011, Garcia2024}. Were such an inflow present, it would lower the current gas-phase metallicity while leaving the integrated stellar enrichment relatively unaffected \citep{Garcia2024}. We note, however, that the currently observed absorber ($\log_{10}(Z/Z_\odot) = -1.05$) is not metal-poor enough relative to this ISM to drive further dilution. Rather, the comparable metallicities of the absorber and G2's ISM suggest they trace a shared reservoir of mildly enriched gas, though this comparison is not strongly constraining given the large uncertainties in $Z_\text{gas}$ for G2 \citep{Ribaudo2011, Wotta_2016}. The \texttt{PROSPECTOR} star-formation histories further indicate prolonged star-forming activity over the past few hundred Myr, consistent with continued replenishment of the ISM by external, metal-poor gas. Together, these results are consistent with a scenario in which periodic accretion of low-metallicity material has contributed to the chemical evolution and star formation history of the galaxy pair, though the available data do not uniquely distinguish this from a tidal or intragroup gas origin within the shared halo environment.

In addition, the low metallicity of the absorber does not favor an outflow origin driven by stellar feedback; instead suggests an origin in accreting intergalactic material that has been moderately enriched by prior star formation activity. In cosmological simulations, it is found that across all halo masses and redshifts considered, IGM accretion provides $\approx  60-80$\% of the CGM gas mass at $\approx R_\text{vir}$, whereas it is only in the inner CGM ($0.1R_\text{vir} \lesssim R \lesssim 0.5R_\text{vir}$) we can expect to find recycled wind material. The wind recycling process is typically concentrated around the scale radius of the halo, leading to the presence of a recycling zone in the inner CGM scaling with the stellar and inner halo structure of galaxies \citep{AnglesAlcazar2017}. The absorber may still trace relic wind material from a more active phase of stellar feedback in either galaxy's history, which has subsequently mixed and become diluted in metallicity within the overlapping halos of the two galaxies. At higher redshifts ($z \gtrsim 1$), where cosmic star formation peaks, large-scale outflows can enrich much larger volumes that extend to, and in some cases beyond, the virial radius, where they get mixed with the adjoining intergalactic gas filaments linked to the large-scale structure. Such material can later migrate back as accreting material, contributing to the recycled gas reservoir at $z \lesssim 1$ \citep[e.g.,][]{Oppenheimer2008, Muratov2015}.

\section{SUMMARY AND CONCLUSION}
\label{sec:summary}

We have presented the analysis of a pLLS at $z_{abs} = 0.87641$ and its connection to a pair of galaxies proximate to the absorber. By combining high-resolution FUV data from $HST$/COS with integral field spectroscopy from $VLT$/MUSE, we have characterized the physical properties of both the absorbing gas and its galactic environment. Our main findings are as follows:

\begin{enumerate}
     \item The absorber is detected in {\ion{H}{1}} from Ly$-\epsilon$ to higher orders, along with {\ion{O}{3}}, {\ion{O}{4}}, {\ion{O}{5}}, {\ion{S}{4}}, and {\ion{S}{5}}. Prominent non-detections include lines of {\ion{C}{2}}, {\ion{N}{2}}, {\ion{O}{2}}, {\ion{Ne}{8}}, and {\ion{S}{6}}. At the resolution of COS, the lines exhibit simple kinematics, well fitted by a single velocity component.

     \item The pLLS has an \ion{H}{1} column density of $\log_{10}[N(\text{\ion{H}{1}})/\text{cm}^{-2}] = 16.03 \pm 0.04$. The line widths of the associated metal ions show significant non-thermal broadening suggestive of turbulent gas flows, or small-scale bulk motions, within the absorbing region. Photoionization modeling constrains the absorber metallicity to $\log_{10}(Z/Z_\odot) = -1.05 \pm 0.05$. 
    
    \item The MUSE data shows two galaxies, G1 and G2, at $|\Delta{v}|$ = 18 and 99~{\kms}, and $\rho/R_\text{vir} \approx 0.9$ from the absorber. The two galaxies themselves have a projected physical separation of $22.6$ kpc, which implies that the galaxies reside within overlapping virial halos. The spectra of both galaxies show prominent nebular emission features, consistent with active star formation.
    
    \item Star formation rates derived from emission-line modeling place G2 on the star-forming main sequence, while G1 exhibits elevated star formation activity ($\approx 2.1\sigma$ above the MS). The high star formation surface densities ($\Sigma_{\text{SFR}} \approx 0.07-0.10\,\mathrm{M}_\odot\,\mathrm{yr}^{-1}\,\mathrm{kpc}^{-2}$) suggest that galaxies are capable of driving galactic winds, though the absorber's properties favor an accretion origin.
    
    \item The quasar sightline is aligned with the projected major axes of both galaxies, at small azimuthal angles ($\alpha \approx 5^\circ - 9^\circ$), but at the edge of their overlapping virial halos. The absorber's velocity offsets from each galaxy are inconsistent with a straightforward continuation of their disk rotation, suggesting that it traces a kinematically distinct, gas component, such as an inflow from the surrounding intergalactic medium. 
    
    \item The absorber's sub-solar metallicity ($\log_{10}(Z/Z_\odot) = -1.05$, above the $z < 1$ pLLS population median of $\log_{10}(Z/Z_\odot) = -1.3$; \citealt{Lehner2019, Wotta2019}), its photoionization temperature ($T \lesssim T_\mathrm{max} \approx 5 \times 10^4$~K, well below $T_\mathrm{vir} \approx 8.5 \times 10^5$~K), and its major-axis alignment are consistent with cool inflowing gas. However, these diagnostics are not unique, and the data are equally consistent with recycled accretion or gas arising from galaxy--galaxy interactions within the shared halo environment.
    
\end{enumerate}

\begin{acknowledgments}
S.R.B. acknowledges the One Nation One Subscription (ONOS) initiative for literature access. S.R.B. also acknowledges the Indian Institute of Space Science and Technology (IIST) for financial support through the PhD institute fellowship. S.C. gratefully acknowledges support from the European Research Council (ERC) under the European Union's Horizon 2020 Research and Innovation program grant agreement No 864361. We thank the anonymous referees for their valuable inputs. The $HST$/COS observations analyzed in this work were obtained from the Mikulski Archive for Space Telescopes (MAST) at the Space Telescope Science Institute. The specific observations can be accessed via \dataset[doi:10.17909/4p1r-jp15]{https://doi.org/10.17909/4p1r-jp15}. This work made use of \textsc{cloudy} \citep{cloudy}, \texttt{VoigtFit} \citep{voigtfit}, \texttt{emcee} \citep{emcee_2013}, \texttt{PROSPECTOR} \citep{Johnson2021}, \texttt{Astropy} \citep{Astropy2013, Astropy2018}, \texttt{NumPy} \citep{Harris2020}, \texttt{Matplotlib} \citep{Hunter2007}, and \texttt{SExtractor} \citep{Bertin1996}.
\end{acknowledgments}

\bibliographystyle{aasjournal}
\bibliography{references_cleaned}

\appendix

\makeatletter
\renewcommand{\thefigure}{A\@arabic\c@figure}
\setcounter{figure}{0}
\renewcommand{\thetable}{A\@arabic\c@table}
\setcounter{table}{0}
\makeatother

\section{Estimating the Pressure of the Hot Halo Gas}
\label{sec:appendixA}
\noindent To assess whether the absorber could be in pressure equilibrium with the hot circumgalactic medium, we estimate the thermal pressure of a hypothetical hot gas halo linked with either of the galaxies in the galaxy pair. We assume that the hot gas is distributed according to the spherical $\beta-$model, a standard parametric density profile of the following form given by \citet{Miller2013}. 
\[
 n(r)~ =~ n_0~\left[1+\left(\frac{r}{r_c}\right)^2\right]^{-1.5\beta}
\]
where $n_0$ is the core number density (number density in the central regions of the halo), $r_c$ is the core radius, and $\beta$ is the slope parameter

\noindent We adopt typical values for these parameters based on observational and simulation work: a core radius $r_c = 0.1 R_\text{vir}$ and a slope $\beta = 0.5$.

\noindent The total baryonic mass present as virialized hot gas ($T \approx 10^6$~K), $M_{\rm hot}$, constitutes only a fraction of the total halo mass ($M_\text{h}$). This is given by 
\[
 M_{\rm hot} = f_{hot}~f_b~M_\text{h}
\]

\noindent where $f_b = 0.16$ is the cosmic baryon fraction \citep{Planck_Collab_2020}, and $f_{hot}$ is the fraction of baryons in the hot phase. This fraction is not well constrained. Based on simulations and observations, we consider a plausible range of $f_{hot}~=~0.1-0.5$ \citep{Stern2016, Pandya2021}. For halo mass, and virial radius, we adopt $M_\text{h}~\approx~7.5\times10^{11}M_\odot$, and $R_\text{vir} = 137$~kpc, an average of both galaxies. 

\noindent The mass of the halo ($M_\mathrm{hot}$) is given by:
\[
M_\mathrm{hot}~=~4\pi \mu m_p n_0\int_0^{R_\mathrm{vir}} \left[ 1+\left(\frac{r}{r_c}\right)^2 \right]^{-1.5\beta}~r^2~dr
\]

\noindent For a fully ionized plasma with a primordial abundance of Hydrogen and Helium, $\mu~\approx~0.6$. 

\noindent For our mean halo properties, this procedure yields a central density $n_0~\approx~(1.3-6.3)\times10^{-3}~\mathrm{cm}^{-3}$. At the absorber's location ($r=0.9~R_\mathrm{vir}$), the density is suppressed by a factor of $\approx~27$, giving a gas density of:
\[
n(r_\mathrm{abs})~\approx~(4.6-23.2)~\times~10^{-5}~\mathrm{cm}^{-3}
\]

\noindent The hot gas is expected to be at the virial temperature of the halo. For a halo of this mass, the virial temperature is $T_\mathrm{vir}~\approx~8.5\times10^5$~K. The corresponding thermal pressure is:
\[ P/k~\approx~(39-197) ~\mathrm{K~cm}^{-3} \]

\noindent This estimated pressure range for the hot halo is comparable to the pressure of the absorbing gas, $P/k~\approx~40~\mathrm{K~cm}^{-3}$, derived from our photoionization modeling (Section~\ref{sec:ionization_modeling}), suggesting that the cool, dense absorber can be in pressure equilibrium with a surrounding hot, diffuse medium. The pLLS might be tracing a cooler and denser gas structure confined by a hot circumgalactic halo such as in cold inflowing streams. 

\clearpage
\clearpage
\begin{table*}
  \centering
  \scriptsize
  \setlength{\tabcolsep}{3.5pt}
  \renewcommand{\arraystretch}{1.5}
  \caption{Equivalent widths, kinematic properties, and column densities of absorption lines at $z=0.87641$. All measurements are derived from the apparent optical depth (AOD) method. The value for the saturated \ion{H}{1} 949 line is a lower limit. $3\sigma$ upper limits are provided for non-detected lines, with those marked as (conta.) known to be blended or potentially contaminated. Velocity windows are defined relative to the systemic redshift of G1 ($z=0.87630$).}
  \label{tab:absorption_lines_updated}

  \begin{tabular}{lrrrrc}
    \hline
    \hline
    {Line} & {Total $W_r$} & {[v$_\text{min}$, v$_\text{max}$]} & {$\Delta v_{90}$} & {$W_r$ in $\Delta v_{90}$} & {log$_{10}$[N$_a$/cm$^{-2}$]} \\
     & (m\AA) & (km s$^{-1}$) & (km s$^{-1}$) & (m\AA) &  \\
    \hline
    \ion{H}{1} 949 & $608 \pm 27$ & [$-30$, 160] & 107 & $447 \pm 17$ & $> 15.92$ \\
    \ion{H}{1} 937 & $434 \pm 21$ & [$-30$, 160] & 108 & $364 \pm 14$ & $15.90 \pm 0.03$ \\
    \ion{H}{1} 930 & $286 \pm 13$ & [$-15$, 95] & 63 & $230 \pm 7$ & $15.84 \pm 0.02$ \\
    \ion{H}{1} 923 & $164 \pm 12$ & [10, 100] & 56 & $140 \pm 9$ & $15.86 \pm 0.02$ \\
    \ion{H}{1} 920 & $126 \pm 11$ & [15, 85] & 40 & $98 \pm 6$ & $15.90 \pm 0.03$ \\
    \ion{H}{1} 919 & $92 \pm 11$ & [15, 80] & 49 & $74 \pm 9$ & $15.79 \pm 0.04$ \\
    \ion{H}{1} 918 & $58 \pm 9$ & [30, 80] & 31 & $47 \pm 7$ & $15.70 \pm 0.05$ \\
    \ion{H}{1} 917 & $59 \pm 9$ & [30, 75] & 33 & $47 \pm 7$ & $15.87 \pm 0.04$ \\
    \ion{O}{3} 832 & $190 \pm 7$ & [5, 105] & 47 & $152 \pm 4$ & $14.43 \pm 0.01$ \\
    \ion{O}{3} 702 & $176 \pm 10$ & [5, 105] & 48 & $119 \pm 4$ & $14.46 \pm 0.02$ \\
    \ion{O}{4} 787 & $231 \pm 7$ & [$-20$, 110] & 74 & $192 \pm 5$ & $14.51 \pm 0.01$ \\
    \ion{O}{4} 608 & $144 \pm 18$ & [$-20$, 110] & 78 & $124 \pm 13$ & $14.67 \pm 0.03$ \\
    \ion{O}{5} 629 & $228 \pm 10$ & [$-20$, 125] & 88 & $191 \pm 7$ & $14.04 \pm 0.01$ \\
    \ion{S}{4} 748 & $72 \pm 11$ & [$-20$, 110] & 81 & $70 \pm 9$ & $13.25 \pm 0.04$ \\
    \ion{S}{5} 786 & $147 \pm 8$ & [$-20$, 125] & 83 & $129 \pm 6$ & $13.12 \pm 0.01$ (conta.) \\
    \hline
    \hline
    \multicolumn{6}{c}{Non-detections ($3\sigma$ upper limits)} \\
    \hline
    \ion{C}{2} 903.9 & $< 55.24$ & [$-25$, 125] & -- & -- & $< 13.00$ \\
    \ion{C}{2} 687 & $< 110^{a}$ & [$-25$, 125] & -- & -- & $< 14.11$ \\
    \ion{C}{2} 903.6 & $< 55.31$ & [$-25$, 125] & -- & -- & $< 13.30$ \\
    \ion{N}{2} 916 & $< 60.92$ & [$-25$, 125] & -- & -- & $< 13.39$ \\
    \ion{N}{3} 685 & -- & [$-25$, 125] & -- & -- & $< 13.40$ (conta.) \\
    \ion{N}{3} 764 & $< 45$ & [$-25$, 125] & -- & -- & $< 13.50$ \\
    \ion{N}{4} 765 & $< 45$ & [$-25$, 125] & -- & -- & $< 12.96$ \\
    \ion{Ne}{8} 770 & $< 51.74$ & [$-25$, 125] & -- & -- & $< 13.58$ \\
    \ion{Ne}{8} 780 & $< 51.55$ & [$-25$, 125] & -- & -- & $< 13.86$ (conta.) \\
    \ion{O}{2} 834 & $< 51.50$ & [$-25$, 125] & -- & -- & $< 13.42$ \\
    \ion{S}{6} 933 & $< 67.41$ & [$-25$, 125] & -- & -- & $< 12.94$ \\
    \ion{S}{6} 945 & $< 54.38$ & [$-25$, 125] & -- & -- & $< 13.25$ (conta.) \\
    \hline
    \hline
    \multicolumn{6}{l}{\scriptsize $^{a}$ Contaminated by Ly$\alpha$ at $z=0.060568$ and other Lyman lines \citep{Tejos}.}
  \end{tabular}
  \renewcommand{\arraystretch}{1.0}
\end{table*}

\begin{figure}
    \centering
    \includegraphics[width=0.9\linewidth]{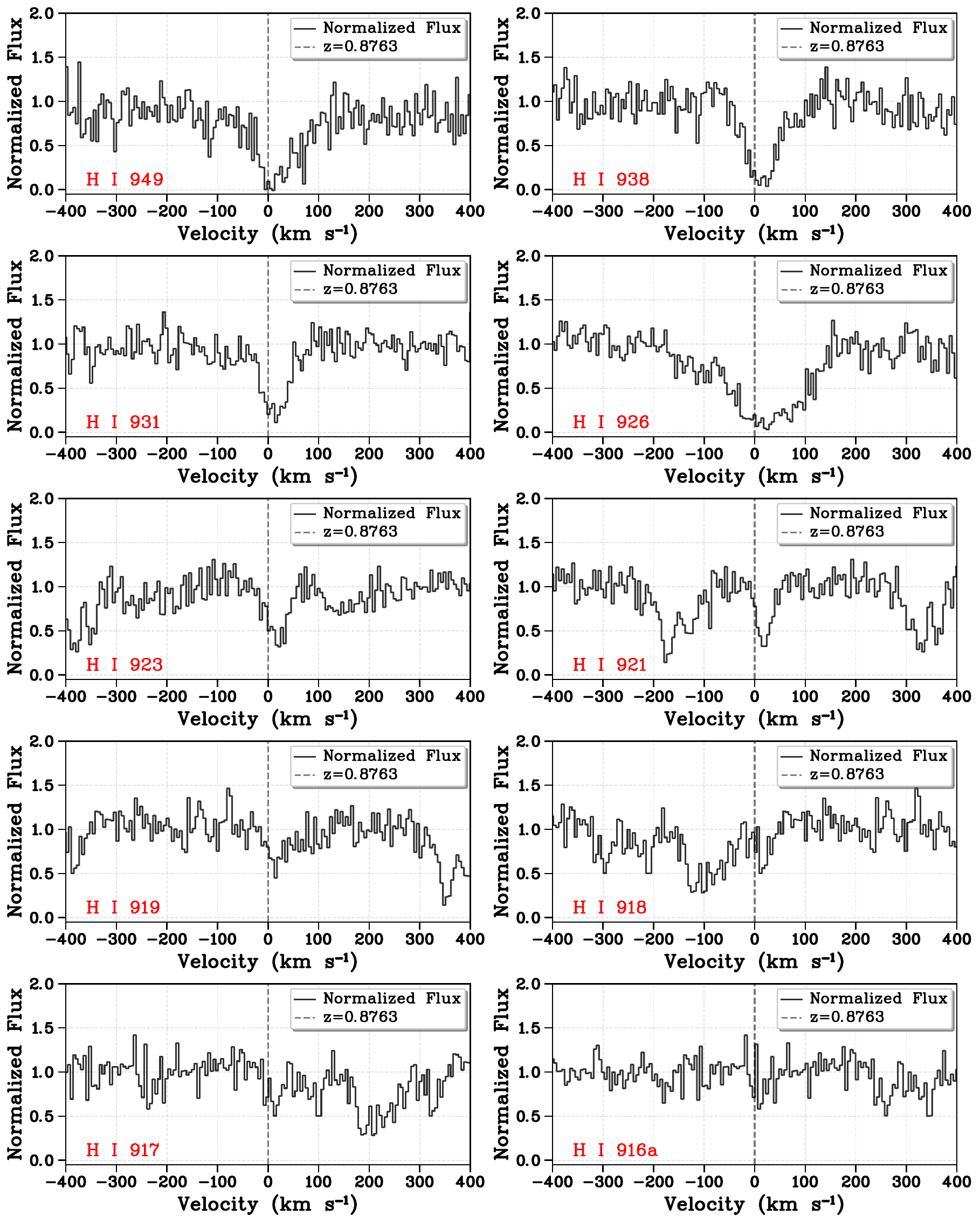}
    \caption{Full velocity plot of the $z = 0.87641$ absorption system (Page 1 of 4). Each panel shows the normalized $HST$/COS flux (black histogram) for \ion{H}{1} Lyman series transitions, from \ion{H}{1} to \ion{H}{1} 916a. The velocity is centered at $z = 0.87630$ (dashed line)}
    \label{fig:sys_plot_page1} 
\end{figure}
\clearpage 

\begin{figure}
    \centering
    \includegraphics[width=0.9\linewidth]{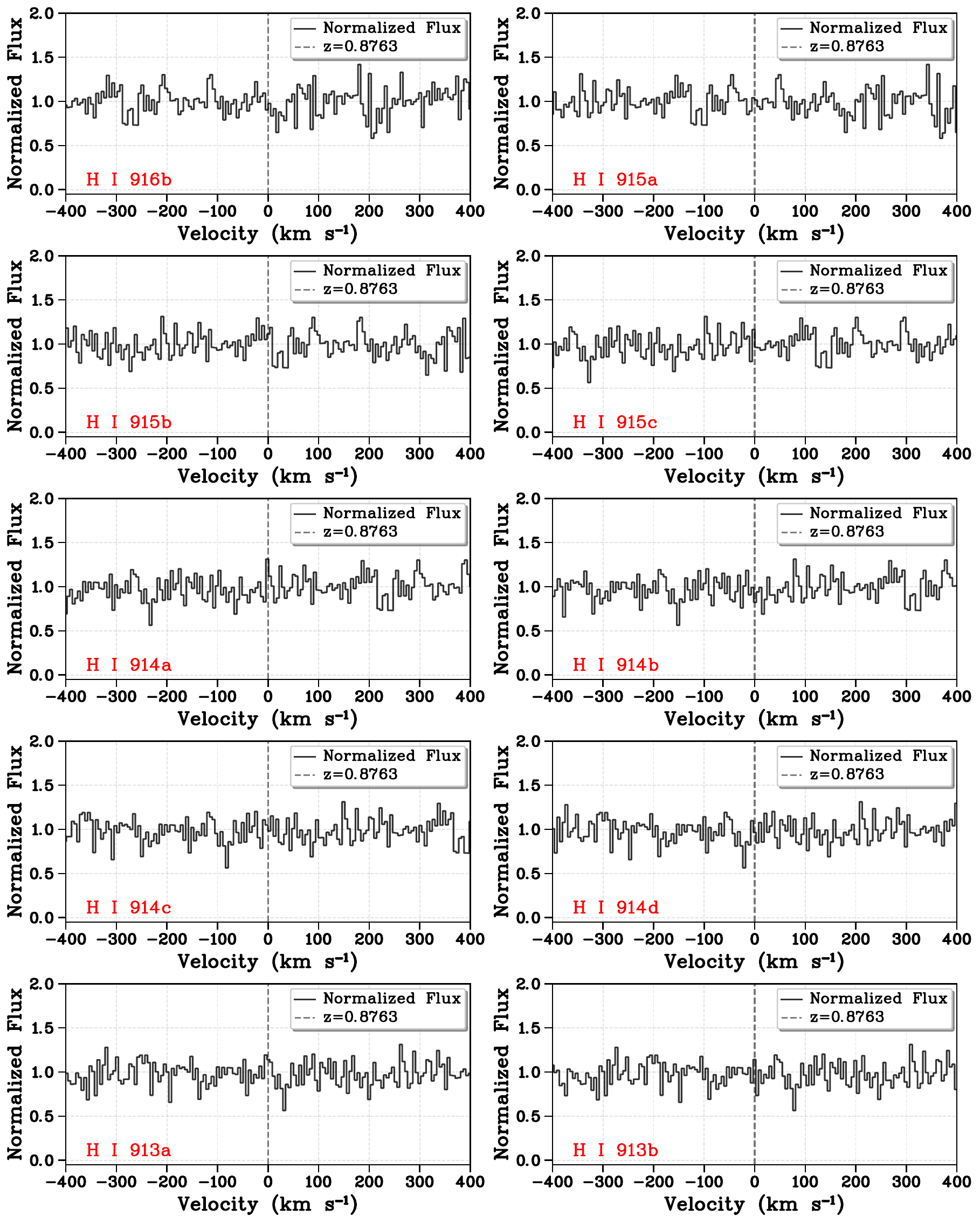}
    \caption{Full velocity plot of the $z = 0.87641$ absorption system (Page 2 of 4). This plot continues the \ion{H}{1} Lyman series transitions, from \ion{H}{1} 916b to \ion{H}{1} 913b}
    \label{fig:sys_plot_page2} 
\end{figure}
\clearpage 

\begin{figure}
    \centering
    \includegraphics[width=0.9\linewidth]{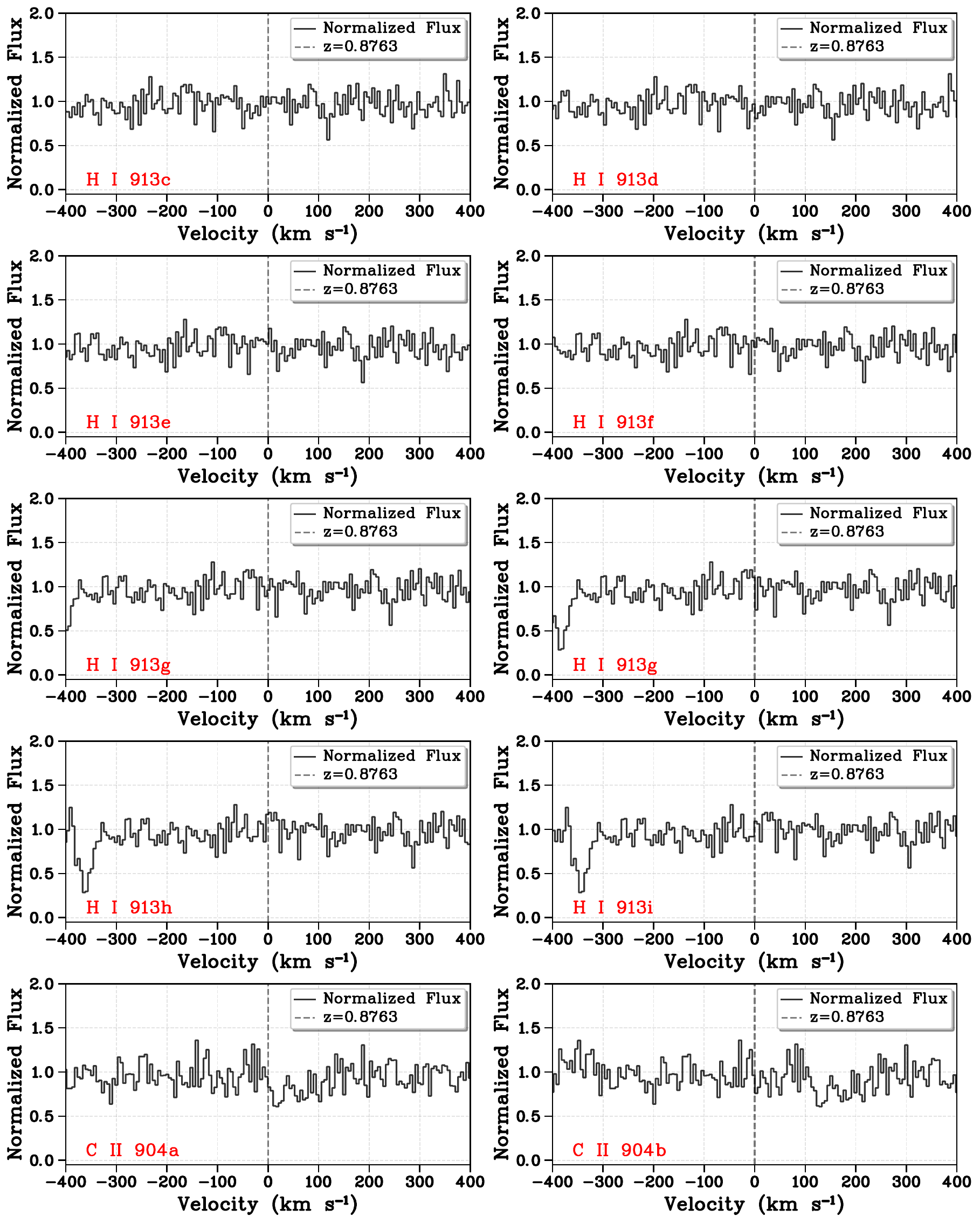}
    \caption{Full velocity plot of the $z = 0.87641$ absorption system (Page 3 of 4). This plot shows the final \ion{H}{1} Lyman series transitions (\ion{H}{1} 913c to \ion{H}{1} 913i) and the non-detected \ion{C}{2} 904 transitions.}
    \label{fig:sys_plot_page3} 
\end{figure}
\clearpage 

\begin{figure}
    \centering
    \includegraphics[width=0.8\linewidth]{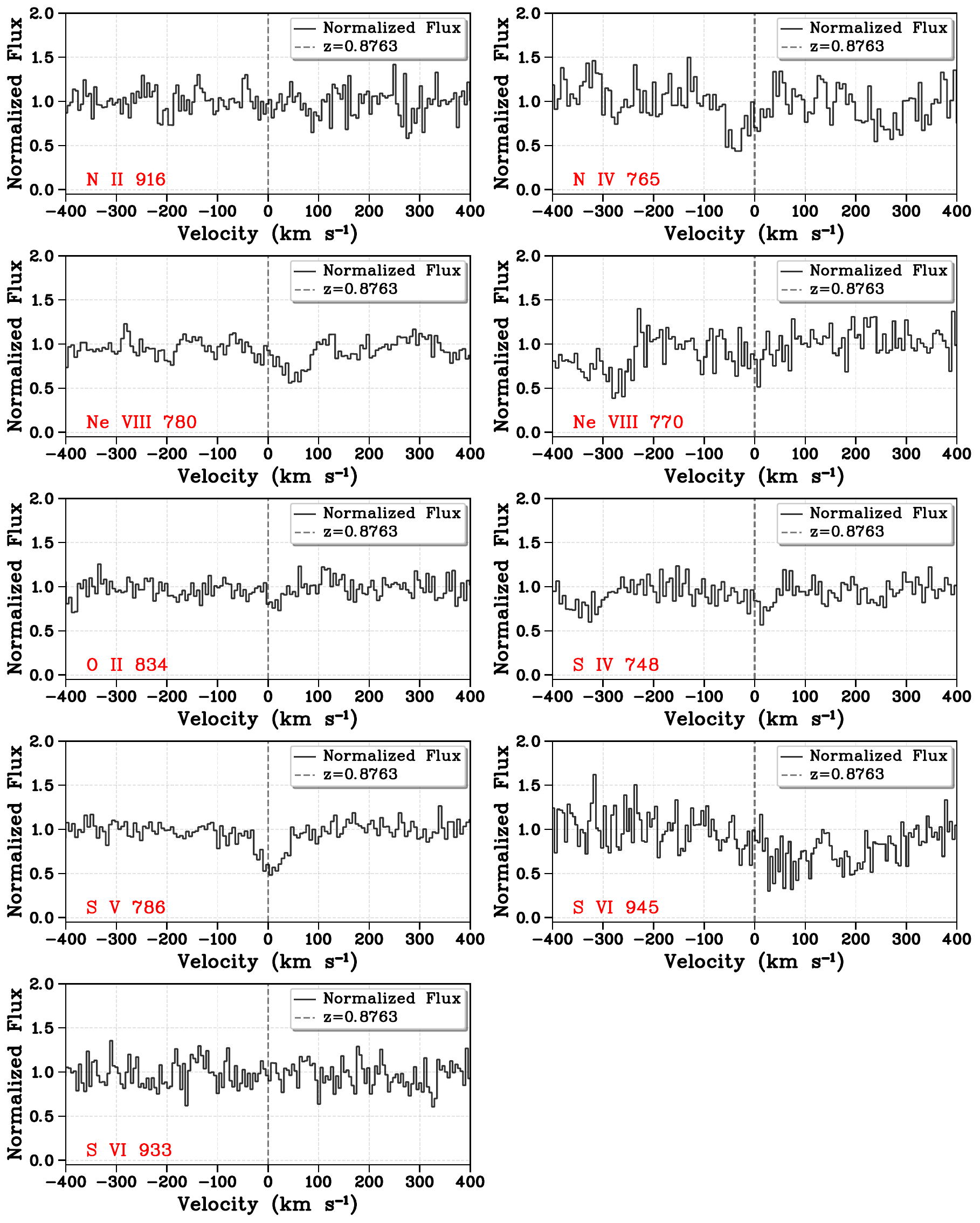}
    \caption{Full velocity plot of the $z = 0.87641$ absorption system (Page 4 of 4). This plot shows various metal ion transitions. \ion{S}{4} 748 and \ion{S}{5} 786 show clear detections, while \ion{N}{2}, \ion{N}{4}, \ion{Ne}{8}, \ion{O}{2}, and \ion{S}{6} are non-detections.}
    \label{fig:sys_plot_page4} 
\end{figure}
\clearpage 

\begin{figure}
    \centering
    \includegraphics[width=0.8\linewidth]{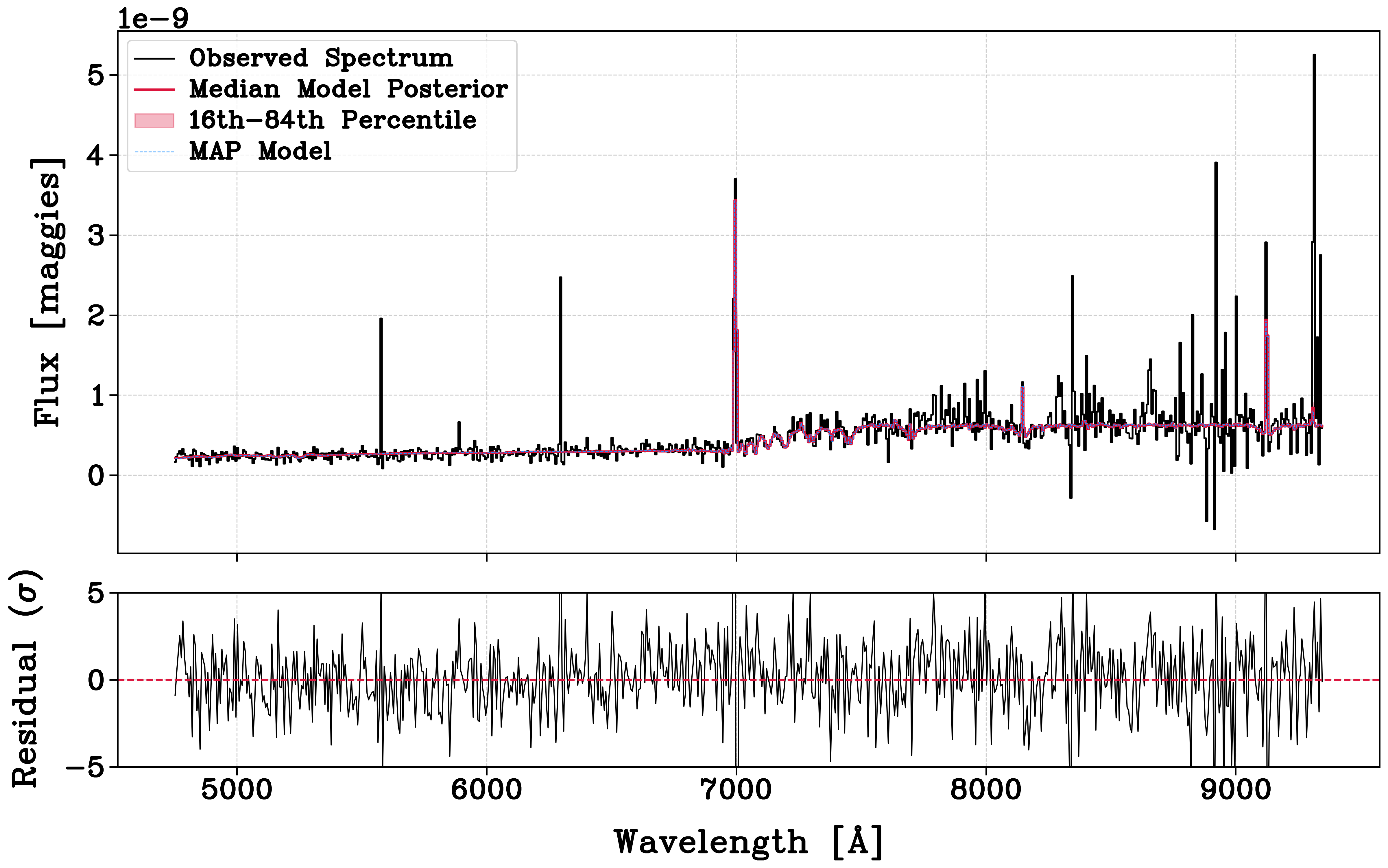}
    \caption{Stellar population synthesis fit for Galaxy G1 using \texttt{PROSPECTOR}. Top panel: The observed MUSE spectrum (black) is shown with the median model posterior (red) and MAP model (blue). Bottom panel: The residual (data $-$ model) normalized by the observational uncertainty ($\sigma$)}
    \label{fig:spectrum_residuals_G1}
\end{figure}

\begin{figure}
    \centering
    \includegraphics[width=0.8\linewidth]{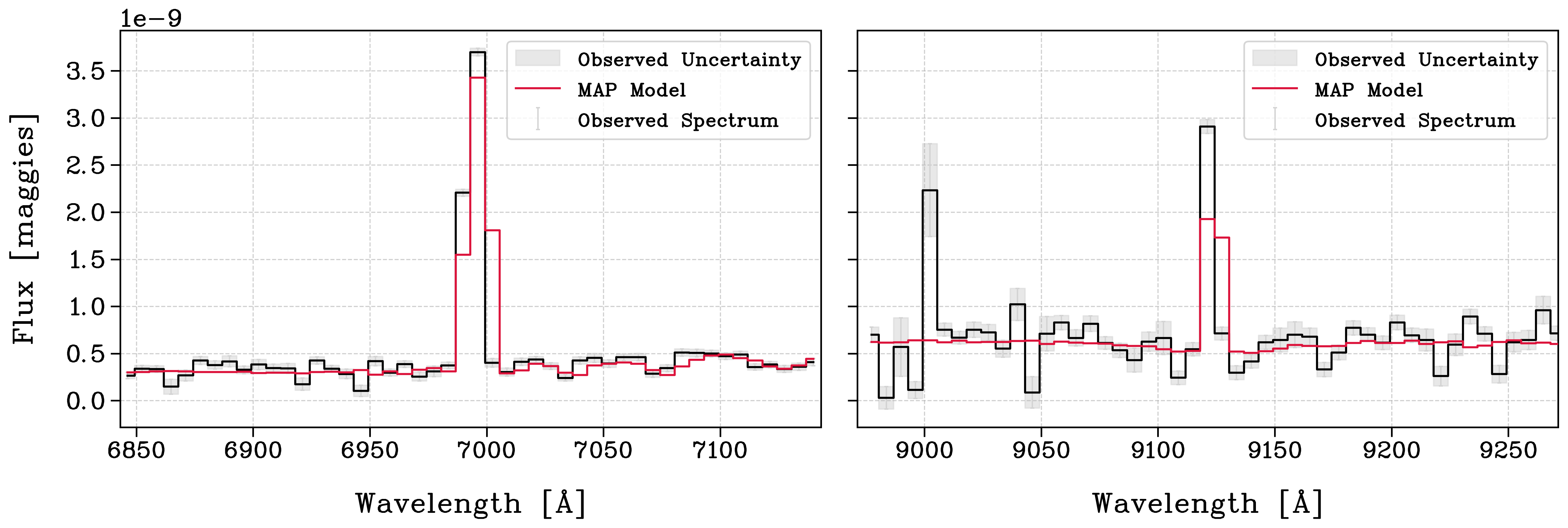}
    \caption{Zoom-in on key emission line fits for Galaxy G1. The observed spectrum (black) and uncertainty (gray) are shown with the best-fit MAP model (red). Left panel: [\ion{O}{2}] $\lambda\lambda 3726, 3729$ doublet. Right panel: H$\beta$ $\lambda 4861$ line.}
    \label{fig:emission_lines_subplots_G1}
\end{figure}

\begin{figure}
    \centering
    \includegraphics[width=0.8\linewidth]{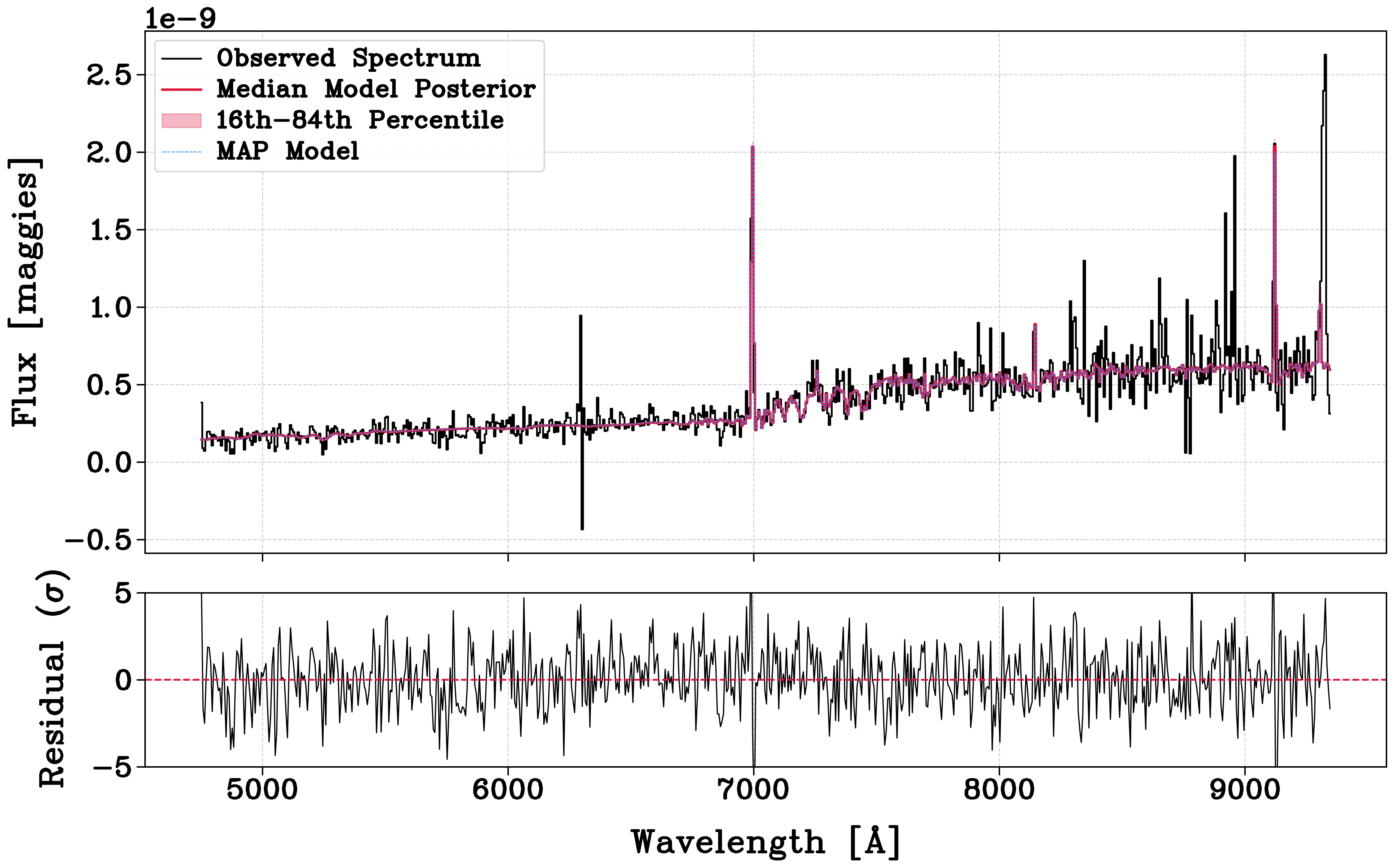}
    \caption{Stellar population synthesis fit for Galaxy G2 using \texttt{PROSPECTOR}. Top panel: The observed MUSE spectrum (black) is shown with the median model posterior (red) and MAP model (blue). Bottom panel: The residual (data $-$ model) normalized by the observational uncertainty ($\sigma$)}
    \label{fig:spectrum_residuals_G2}
\end{figure}

\begin{figure}
    \centering
    \includegraphics[width=0.8\linewidth]{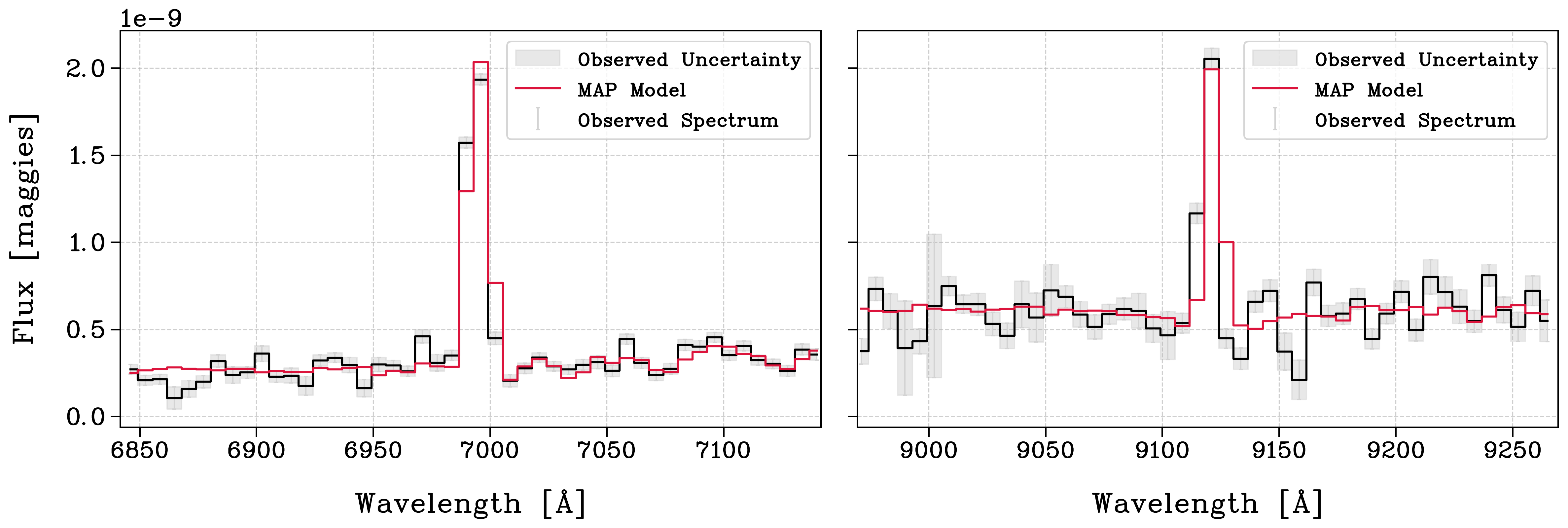}
    \caption{Zoom-in on key emission line fits for Galaxy G2. The observed spectrum (black) and uncertainty (gray) are shown with the best-fit MAP model (red). Left panel: [\ion{O}{2}] $\lambda\lambda 3726, 3729$ doublet. Right panel: H$\beta$ $\lambda 4861$ line}
    \label{fig:emission_lines_subplots_G2}
\end{figure}


\begin{figure}
\centering

\begin{minipage}[t]{0.49\linewidth}
    \centering
    \includegraphics[width=\linewidth]{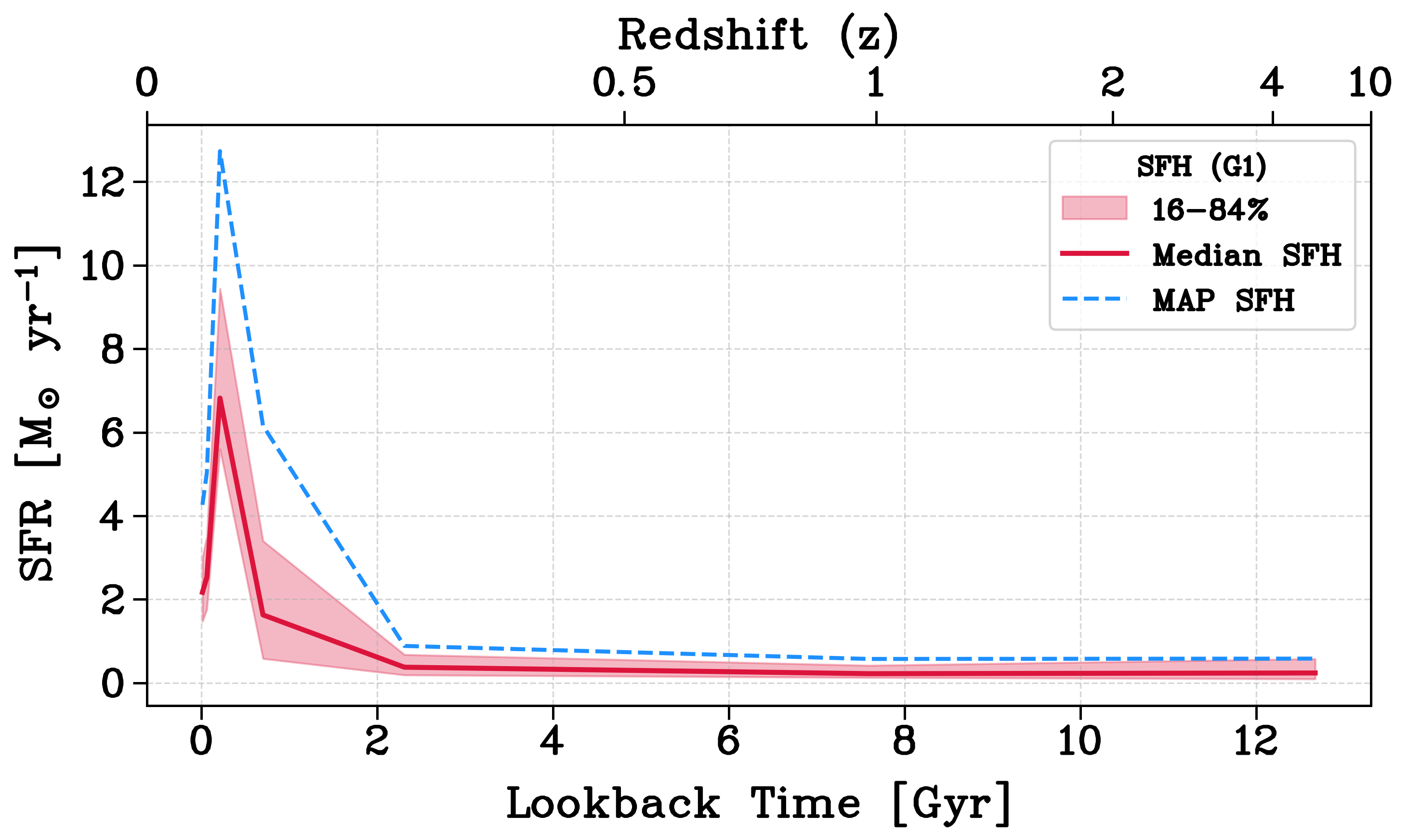}
    \label{fig:SFH_fit_1}
\end{minipage}
\hfill
\begin{minipage}[t]{0.49\linewidth}
    \centering
    \includegraphics[width=\linewidth]{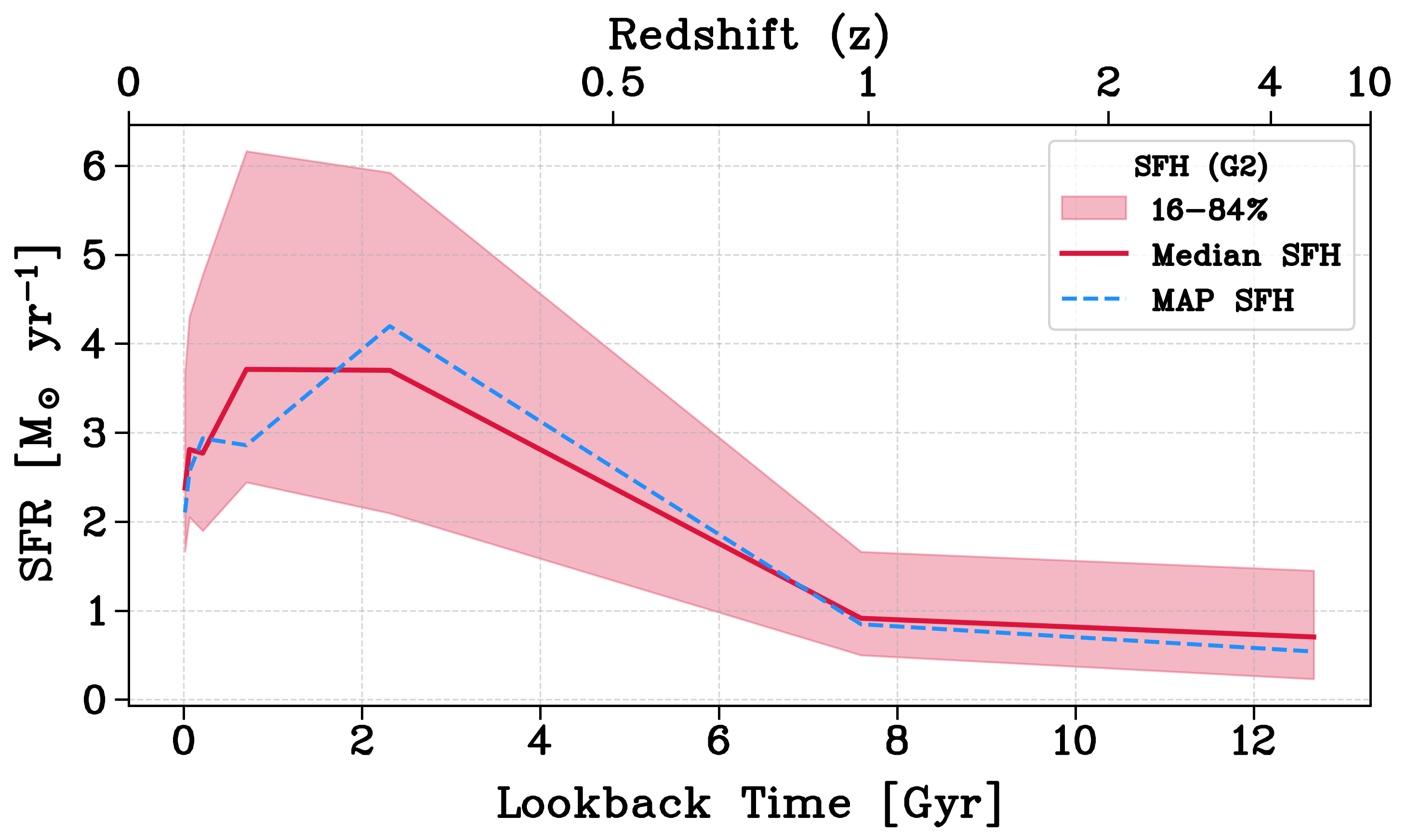}
    \label{fig:SFH_fit_2}
\end{minipage}

\caption{Non-parametric star formation histories (SFH) derived from the \texttt{PROSPECTOR} analysis. Left panel: The SFH for Galaxy G1. Right panel: The SFH for Galaxy G2. In both plots, the solid red line shows the median SFH from the posterior, the pink shaded area represents the 16th$-$84th percentile uncertainty, and the dashed blue line is the SFH from the MAP parameters.}
\label{fig:SFH}
\end{figure}

\clearpage


\end{document}